\documentclass[useAMS,usedcolumn,usenatbib,usegraphicx,sort]{mn2e}

\usepackage{epsfig,graphicx,natbib}
\usepackage{./reference/mycite}
\usepackage{amssymb}
\usepackage{gensymb}
\usepackage{amsfonts}
\usepackage{amsmath}
\usepackage{color}
\usepackage{lscape,longtable}
\usepackage{times}
 
\begin{document}

\title[DES+VHS]{Combining Dark Energy Survey Science Verification Data with Near Infrared Data from the ESO VISTA Hemisphere Survey} 
\author[M. Banerji et al.]{ \parbox{\textwidth}
{Manda Banerji$^{1}$\thanks{E-mail: m.banerji@ucl.ac.uk}, S. Jouvel$^{1}$, H. Lin$^{2}$, R. G. McMahon$^{3,4}$, O. Lahav$^{1}$, F.~J.~Castander$^{5}$, F. B. Abdalla$^{1}$, E. Bertin$^{6}$, S. E. Bosman$^{3}$, A. Carnero$^{7,8}$, M. Carrasco Kind$^{9}$, L. N. da Costa$^{7,8}$, D. Gerdes$^{10}$, J. Gschwend$^{7,8}$, M. Lima$^{11,8}$, M. A. G. Maia$^{7,8}$, A. Merson$^{3}$, C. Miller$^{10}$, R. Ogando$^{7,8}$, P. Pellegrini$^{7,8}$, S. Reed$^{3}$, R. Saglia$^{12}$, C. S\'{a}nchez$^{13}$, S. Allam$^{14, 2}$, J. Annis$^{2}$, G. Bernstein$^{15}$, J. Bernstein$^{16}$, R. Bernstein$^{17}$, D. Capozzi$^{18}$, M. Childress$^{19,20}$, Carlos E. Cunha$^{21}$, T. M. Davis$^{22,20}$, D.~L.~DePoy$^{23}$, S. Desai$^{24,25}$, H.~T.~Diehl$^{2}$, P. Doel$^{1}$, J. Findlay$^{3}$, D.~A.~Finley$^{2}$, B. Flaugher$^{2}$, J. Frieman$^{2}$, E. Gaztanaga$^{5}$, K. Glazebrook$^{26}$, C. Gonz\'alez-Fern\'andez$^{3}$, E. Gonzalez-Solares$^{3}$, K. Honscheid$^{27}$, M. J. Irwin$^{3}$, M. J. Jarvis$^{28,29}$, A. Kim$^{30}$, S. Koposov$^{3}$, K. Kuehn$^{31}$, A. Kupcu-Yoldas$^{3}$, D. Lagattuta$^{26,20}$, J. R. Lewis$^{3}$, C. Lidman$^{31}$, M. Makler$^{32}$, J. Marriner$^{2}$, Jennifer L.~Marshall$^{23}$, R. Miquel$^{13,33}$, Joseph J. Mohr$^{24,25,12}$, E. Neilsen$^{2}$, J. Peoples$^{2}$, M. Sako$^{15}$, E. Sanchez$^{34}$, V. Scarpine$^{2}$, R. Schindler$^{35}$, M. Schubnell$^{10}$, I. Sevilla$^{34}$, R. Sharp$^{36}$, M. Soares-Santos$^{2}$, M.~E.~C. Swanson$^{37}$, G. Tarle$^{11}$, J. Thaler$^{38}$, D. Tucker$^{2}$, S. A. Uddin$^{26,20}$, R. Wechsler$^{21}$, W. Wester$^{2}$, F. Yuan$^{38,19}$, J. Zuntz$^{39}$
}}

\maketitle

\begin{abstract} 

We present the combination of optical data from the Science Verification phase of the Dark Energy Survey (DES) with near infrared data from the ESO VISTA Hemisphere Survey (VHS). The deep optical detections from DES are used to extract fluxes and associated errors from the shallower VHS data. Joint 7-band ($grizYJK$) photometric catalogues are produced in a single 3 sq-deg DECam field centred at 02h26m$-$04d36m where the availability of ancillary multi-wavelength photometry and spectroscopy allows us to test the data quality.  Dual photometry increases the number of DES galaxies with measured VHS fluxes by a factor of $\sim$4.5 relative to a simple catalogue level matching and results in a $\sim$1.5 mag increase in the 80\% completeness limit of the NIR data. Almost 70\% of DES sources have useful NIR flux measurements in this initial catalogue. Photometric redshifts are estimated for a subset of galaxies with spectroscopic redshifts and initial results, although currently limited by small number statistics, indicate that the VHS data can help reduce the photometric redshift scatter at both $z<0.5$ and $z>1$. We present example DES+VHS colour selection criteria for high redshift Luminous Red Galaxies (LRGs) at $z\sim0.7$ as well as luminous quasars. Using spectroscopic observations in this field we show that the additional VHS fluxes enable a cleaner selection of both populations with $<$10\% contamination from galactic stars in the case of spectroscopically confirmed quasars and $<0.5\%$ contamination from galactic stars in the case of spectroscopically confirmed LRGs. The combined DES+VHS dataset, which will eventually cover almost 5000 sq-deg, will therefore enable a range of new science and be ideally suited for target selection for future wide-field spectroscopic surveys.

\end{abstract}


\section{INTRODUCTION}

Observational astronomy is currently experiencing a golden age of digital data. The advent of sensitive electronic cameras that can be mounted on large aperture telescopes has enabled a new generation of wide-field ground-based galaxy surveys at optical and near infrared wavelengths. These surveys aim to obtain photometric data for millions of galaxies over thousands of square degrees of sky. Wide-field optical survey astronomy was revolutionised by  data from the Sloan Digital Sky Survey (SDSS; \citealt{York:00}). Data Release 9 of SDSS \citep{Ahn:12} contains more than 930 million catalogued objects over 14,555 sq-deg of the northern hemisphere corresponding to a source density of $\sim$64,000/deg$^2$. The 5000 sq-deg Dark Energy Survey (DES) represents the next leap forward in the southern hemisphere with an expected source density of $>$100,000/deg$^2$. DES is a wide-field optical survey optimised for studies of weak gravitational lensing, large-scale structure and galaxy clusters, in particular from the millimeter wavelength South Pole Telescope (SPT) Sunyaev-Zeldovich (SZ) cluster survey. In addition, DES also includes a deeper multi-epoch supernova survey over 30 sq-deg. The optical DES data are complemented by near infrared (NIR) data from the VISTA Hemisphere Survey (VHS; \citealt{McMahon:13}). Together these two surveys will eventually provide a unique 7-band photometric dataset over $\sim$4500 sq-deg of the southern sky. Figure \ref{fig:area} shows the coverage of all VHS data processed until April 2014 in the DES region, together with the DES Science Verification observations.   

DES differs from previous optical surveys like SDSS due to the red-sensitive CCDs used by the DECam camera and its high throughput out to wavelengths of almost 1$\mu$m in the $z$ and $Y$ filters. Near infrared photometry in the $J$ and $K$-bands out to $\sim$2$\mu$m would therefore form a natural extension to the DES wavelength coverage. The addition of NIR photometry to optical data provides several advantages. Firstly, at $z\gtrsim1.0$, the 4000\AA\@ break in early type galaxies which is commonly used as a photometric redshift discriminator, moves from the optical to the NIR bands. The addition of NIR data may therefore lead to significantly more accurate photometric redshifts for galaxies at $z\gtrsim1$ \citep{Banerji:08, Jarvis:13}. These galaxies will serve as important probes of the evolution of large-scale structure with redshift. NIR data is also extremely sensitive to the reddest galaxies which often reside in galaxy clusters, as well as active galactic nuclei (AGN) at all redshifts and can therefore help with their identification. Many of these sources will be targeted with the next generation of wide-field spectroscopic surveys such as DESI \citep{Levi:13} and 4MOST \citep{deJong:12} and it is important to assess the quality of the photometric data to judge its suitability for target selection (e.g. \citealt{Jouvel:13}). The DES+VHS photometric data will also contribute significantly to the landscape of multi-wavelength surveys in the southern sky over the next decade, providing interesting targets for both the Atacama Large Millimeter Array (ALMA) and the Extremely Large Telescope (E-ELT) and enabling the identification of optical counterparts for the Square Kilometer Array (SKA) pathfinders and eventually leading up to the full SKA. Finally, it should be stressed that the techniques for combining the data outlined here as well as the science applications that are described are just as relevant to combining optical and near infrared data from many other ongoing wide-field galaxy surveys - e.g. VST-ATLAS, VST-KiDS, PanSTARRS, HSC, Skymapper, VISTA VIKING. 

In this paper we describe joint optical+ NIR photometric catalogues from DES + VHS, produced over a single DECam field covering 3 sq-deg of the DES Science Verification data. We use the deeper DES detection images to extract fluxes and associated errors from the shallower NIR data. Section \ref{sec:data} describes the surveys used in this work and Section \ref{sec:method} describes the dual photometry method. The astrometry, photometry and star-galaxy separation of the combined catalogues are checked in Section \ref{sec:astrom}, \ref{sec:photo} and \ref{sec:sg} respectively. While this sort of dual photometry method has been used extensively in astronomical surveys for the construction of multi-wavelength photometric catalogues, it is usually applied to datasets that are comparable in terms of their depth and sensitivity. By contrast, the VHS data is significantly shallower than DES, which means that the majority of DES detections lie below the 5$\sigma$ level of the VHS images. A key aim of this paper is to assess whether enough useful information can be extracted even from low signal-to-noise flux measurements, to enhance the science that will be possible with DES data alone. With this in mind, we also present some illustrative science applications of our combined data including photometric redshift estimates (Section \ref{sec:photoz}) and target selection of both Luminous Red Galaxies and quasars (Section \ref{sec:target}). 
 
All magnitudes are on the AB system. This is the native photometric system for DES and the VHS magnitudes have been converted from Vega to AB using the following corrections: $J_{\rm{AB}}$=J$_{\rm{Vega}}$+0.937, $K_{\rm{AB}}$=K$_{\rm{Vega}}$+1.839. 

\section{SURVEY DATA}

\label{sec:data}

\subsection{Dark Energy Survey (DES)}

\label{sec:des}

The Dark Energy Survey \citep{Frieman:13, Flaugher:05} is an optical survey that is imaging 5000 sq-deg of the southern celestial hemisphere in the $grizY$ bands using a dedicated camera, DECam \citep{Flaugher:12, Diehl:12} on the Blanco 4-m telescope in CTIO. The survey aims to understand the nature and evolution of dark energy using a multi-probe approach, utilising measurements of large-scale structure, weak gravitational lensing, galaxy cluster counts and a dedicated multi-epoch supernova survey over 30 sq-deg. The wide survey will go down to a nominal $i$-band magnitude limit of 25.3 (5$\sigma$; PSF) or 24.0$\pm$0.10 (10$\sigma$; galaxy). The camera obtained its first light images in September 2012 and this was followed by a period of Commissioning and Science Verification observations (SV hereafter) lasting between October 2012 and February 2013. Several high priority fields were targetted as part of this SV phase including well-known spectroscopic training set fields overlapping spectroscopic surveys like VVDS, ACES and zCOSMOS, several well-known galaxy clusters as well as $\sim$250 sq-deg of area covered by the SPT survey. 

DECam imaging overlapping data from deep spectroscopic redshift surveys 
were obtained for the following four fields: SN-X3, SN-C3, 
VVDS F14, and COSMOS \citep{Sanchez:14}. Each of the four fields covers about the area of single 
DECam pointing, or about 3 sq-deg and more details can be found in \citet{Sanchez:14}. All fields include imaging in the 6 filters $ugrizY$.
The data have been processed to two imaging depths: ``main'', corresponding
to approximately DES main survey exposure times, and ``deep'', corresponding 
to about 3 times the exposure of a single visit to a DES supernova 
deep field (for SN-X3 and SN-C3) or deeper (for COSMOS).
Details of the data reduction and the exposure times used are given in \citet{Sanchez:14}. In this paper, we work with the DECam imaging
in the SN-X3 field only, processed to the ``main" survey depth. We only use the DES $grizY$ photometry
as $u$-band data will not be available for the main wide-field survey. The SN-X3 field has been chosen
from among the four spectroscopic training set fields as it is the only one with complete overlap with
VHS near infrared imaging. The SN-C3 field is partially covered by the VISTA VIKING survey, which
complements VHS and the VVDS F14 and COSMOS fields lie outside the VHS survey region. 

The data were processed using the same codes used by DES
Data Management (DESDM) to process DES imaging data \citep{Mohr:12, Desai:12}, in particular 
for image detrending, astrometric calibration (SCAMP; \citealt{Bertin:06}), image remapping and 
coaddition (SWarp v2.34.0; \citealt{Bertin:02})
and object detection and photometry (SExtractor v2.18.7; \citealt{Bertin:96}). The typical photometric calibration uncertainty is $<$5\%. The $r$-band image was used
as the detection image in construction of these early photometric catalogues although we note that $riz$ images are now used by DESDM for object detection in order to provide increased sensitivity to red objects. For the purposes of our analysis however, the $r$-band data is significantly deeper than the NIR VHS data so using the $r$-band for detection does not affect the statistical properties of the joint optical+NIR source population although we note that these catalogues are likely to be incomplete to very rare, extremely red objects.  The data were generally processed by running these codes in 
``standalone'' mode at Fermilab, rather than by running them within the 
DESDM processing framework at NCSA.  Running ``standalone'' was needed as
the DESDM framework was not yet fully setup at the time (Spring 2013)
to process and calibrate the data for these isolated fields all the way 
through to image coaddition.

Though we basically used the DESDM codes, there were some detailed
differences in processing and photometric calibration that are highlighted in
\citet{Sanchez:14}. For the purposes of this paper however, the DECam imaging over
SN-X3 is representative of the optical photometry that will be available from the
final DES survey.

\subsection{VISTA Hemisphere Survey (VHS)}

\label{sec:vhs}

The VISTA Hemisphere Survey (VHS) is a NIR photometric survey being conducted using the VISTA telescope in Chile, that aims to cover 18,000 sq-deg of the southern celestial hemisphere to a depth 30 times fainter than the 2MASS survey in at least two wavebands ($J$ and $K_s$). In the South Galactic Cap, $\sim$4500 sq-deg of sky overlapping DES is being imaged deeper in order to better supplement the optical data available from DES. This area is known as VHS-DES hereafter and goes down to median depths of $J_{\rm{AB}}=21.2$ and $K_{\rm{AB}}=20.4$ for 5$\sigma$ detection of a point source. The remainder of the high galactic latitude sky is being imaged in the $YJK$ bands and combined with optical photometry from the VST-ATLAS survey. This VHS-ATLAS area has 5$\sigma$ nominal magnitude limits of $J=20.9$ and $K=19.8$. Some portions of VHS-DES and VHS-ATLAS also include data in the $H$-band. The low galactic latitude sky is known as VHS-GPS and also goes down to $K<19.8$. In this paper, we work with the VHS-DES images and catalogues overlapping the 3 sq-deg DECam pointing centred on the SN-X3 field.

\begin{figure*}
\begin{center}
\includegraphics[scale=0.5,angle=0]{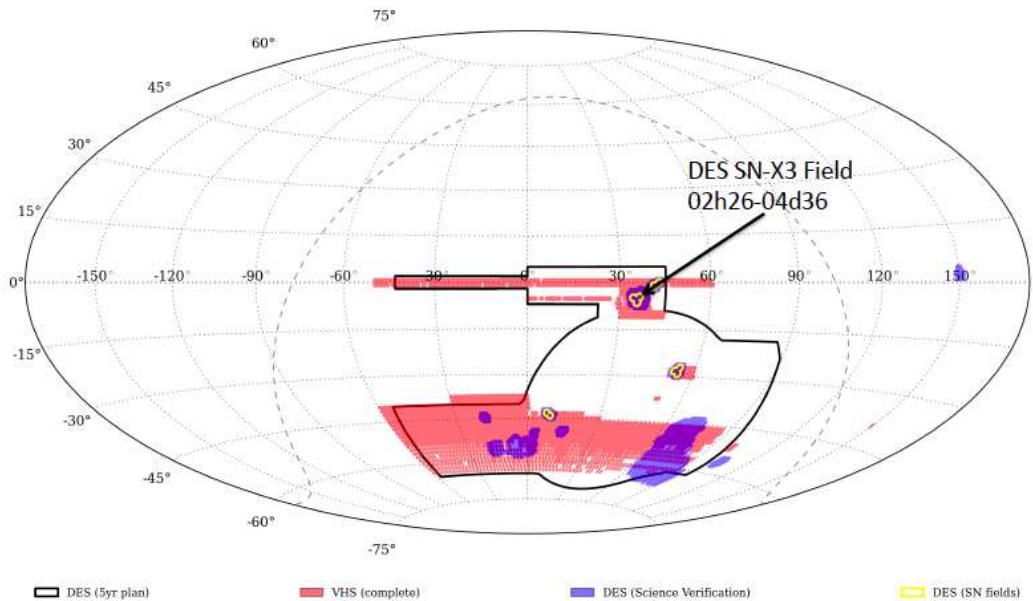}
\caption{The 5yr DES footprint together with DES pointings obtained as part of the DES Science Verification observations (blue) and all VHS pointings overlapping DES and processed until the end of April 2014 (red). The location of all the DES deep supernova fields as well as the DES SN-X3 field used for this study are also marked.}
\label{fig:area}
\end{center}
\end{figure*}

We now introduce several definitions pertinent to the VHS data that will be relevant later. The VIRCAM camera used for VHS imaging constitutes a sparse array of 16 individual detectors that cover a region of 0.595 sq-deg. In order to get contiguous coverage of the 1.5 sq-deg field-of-view (1.02 deg in RA and 1.48 deg in DEC), six exposures are therefore required. These six exposures are termed pawprints and together they produce a single coadd tile. Due to the smaller size of the VIRCAM field of view relative to DECam, each DECam pointing overlaps multiple VIRCAM tiles which are combined as detailed later.  

VHS imaging data has been processed using the VISTA Data Flow System (VDFS; \citealt{Emerson:04, Irwin:04, Hambly:04}) operated by the Cambridge Astronomical Survey Unit (CASU). Data processing steps follow standard procedures for infrared photometry instrumental signature removal including bias, non-linearity, dark, flat and fringing corrections. Sky background tracking and removal are done using all observations executed during a night. The pawprint images are then combined into a coadd tile. Photometric calibration is done using 2MASS on a tile by tile basis and makes use of the 2MASS $JHK_S$ stellar photometry  to calibrate the VHS data as detailed in \citet{Hodgkin:09}. Specifically, the following colour equations are derived between the 2MASS and VISTA $J$ and $K_s$ filters:

\begin{equation}
J_{\rm{VISTA}}=J_{\rm{2MASS}} - 0.077 (J-H)_{\rm{2MASS}} 
\label{eq:Jcal}
\end{equation}

\begin{equation}
K_{\rm{VISTA}}=K_{\rm{2MASS}} + 0.010 (J-K)_{\rm{2MASS}} 
\label{eq:Kcal}
\end{equation}

The photometric calibration in the $J$ and $K_s$-bands has a typical uncertainty of $<$1.5\% (Hodgkin et al. in preparation). Finally, catalogues are generated using the astrometry, photometry, shape and Data Quality Control information. All images and catalogues used in this work are supplied with the appropriate astrometric and photometric calibrations from VDFS. 

\subsection{Spectroscopic Catalogue}

\label{sec:ozdes}

Spectroscopic follow-up of the DES Supernova Fields (including the SN-X3 field), is currently being conducted as part of the OzDES survey on the Anglo-Australian Telescope (Yuan et al. in preparation). The fibre fed spectrograph on the AAT allows OzDES to pursue several parallel scientific goals. A strong focus is on spectroscopic follow-up of supernova host galaxies and repeat observations of Active Galactic Nuclei for reverberation mapping. Other targets include galaxy clusters, Luminous Red Galaxies, Emission Line Galaxies, as well as a flux-limited sample of photometric redshift calibration targets. OzDES is a 5 year survey which began in 2013 and the first season of observations is now complete. OzDES redshifts have been combined with other redshifts in these fields from spectroscopic surveys including SDSS \citep{Eisenstein:01, Strauss:02, Ahn:12}, 6dF \citep{Jones:09}, GAMA \citep{Driver:11, Hopkins:13}, VVDS \citep{LeFevre:05}, VIPERS \citep{Garilli:13} and SNLS supernova hosts \citep{Lidman:13}. 

We make use of this master spectroscopic redshift catalogue\footnote{http://des-docdb.fnal.gov:8080/cgi-bin/ShowDocument?docid=8084} to assess the quality and utility of the DES+VHS catalogues in much of the analysis that follows but note here that this does not form a homogenous population of spectroscopically confirmed galaxies. The GAMA and 6dF surveys in particular constitute the brightest and lowest redshift galaxies in the local Universe while deeper surveys like VVDS and VIPERS extend to much higher redshifts. The OzDES spectroscopically confirmed galaxies too do not form a uniform flux-limited sample due to the fact that at a fixed magnitude, redshifts are more easily determinable for certain sub-classes of galaxies in the OzDES survey. These caveats should be kept in mind when photometric redshifts for this catalogue are presented in Section \ref{sec:photoz}. 

\section{METHOD}

\label{sec:method}

\begin{figure*}
\begin{center}
\begin{tabular}{ccc}
\large{DES $r$ (260 sources)} & \large{VHS $J$ S/N$>$10 (51 sources)} &   \large{VHS $J$ S/N$>$5 (84 sources)}\\
\includegraphics[scale=0.4,angle=0]{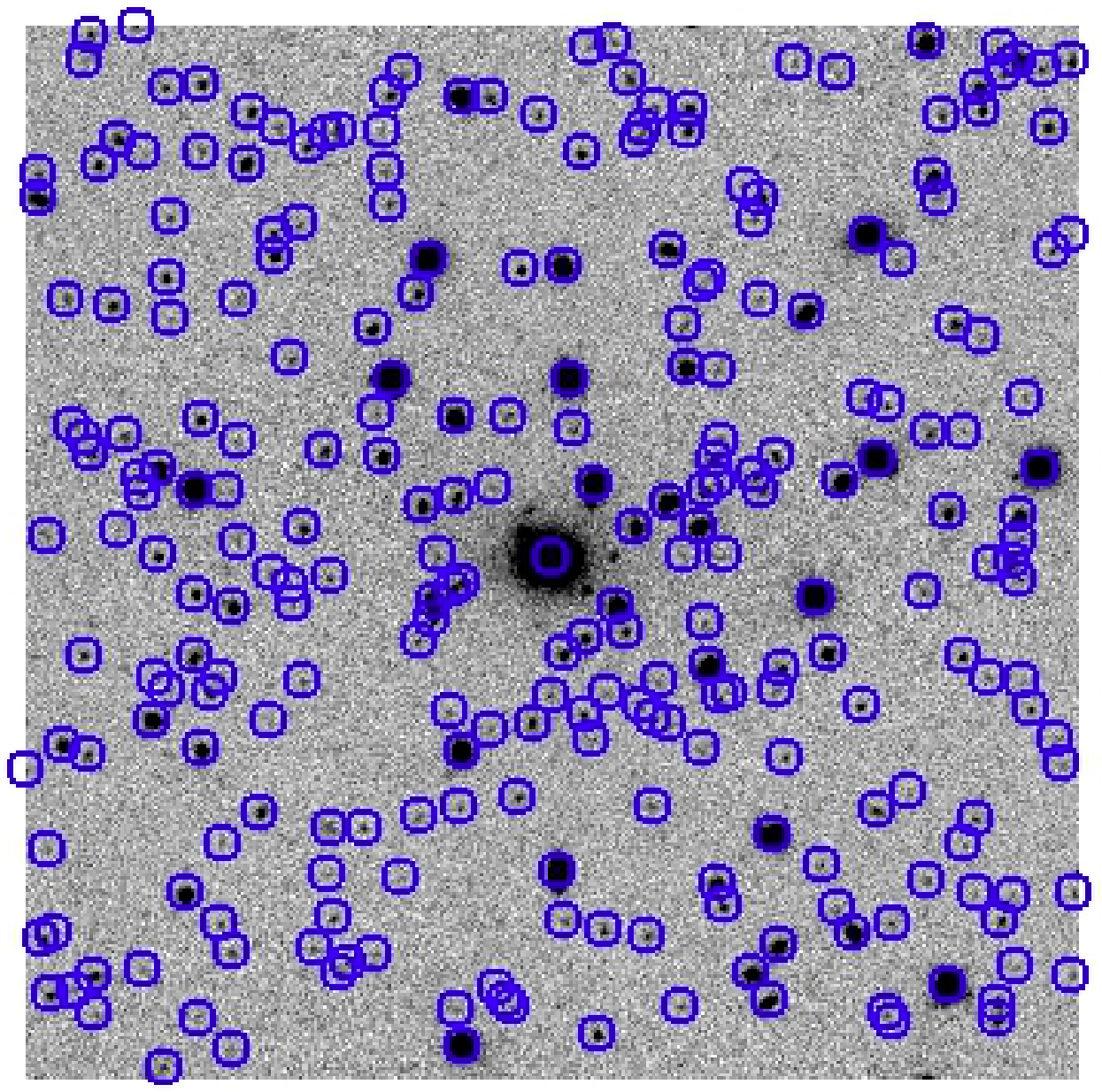} & \includegraphics[scale=0.4,angle=0]{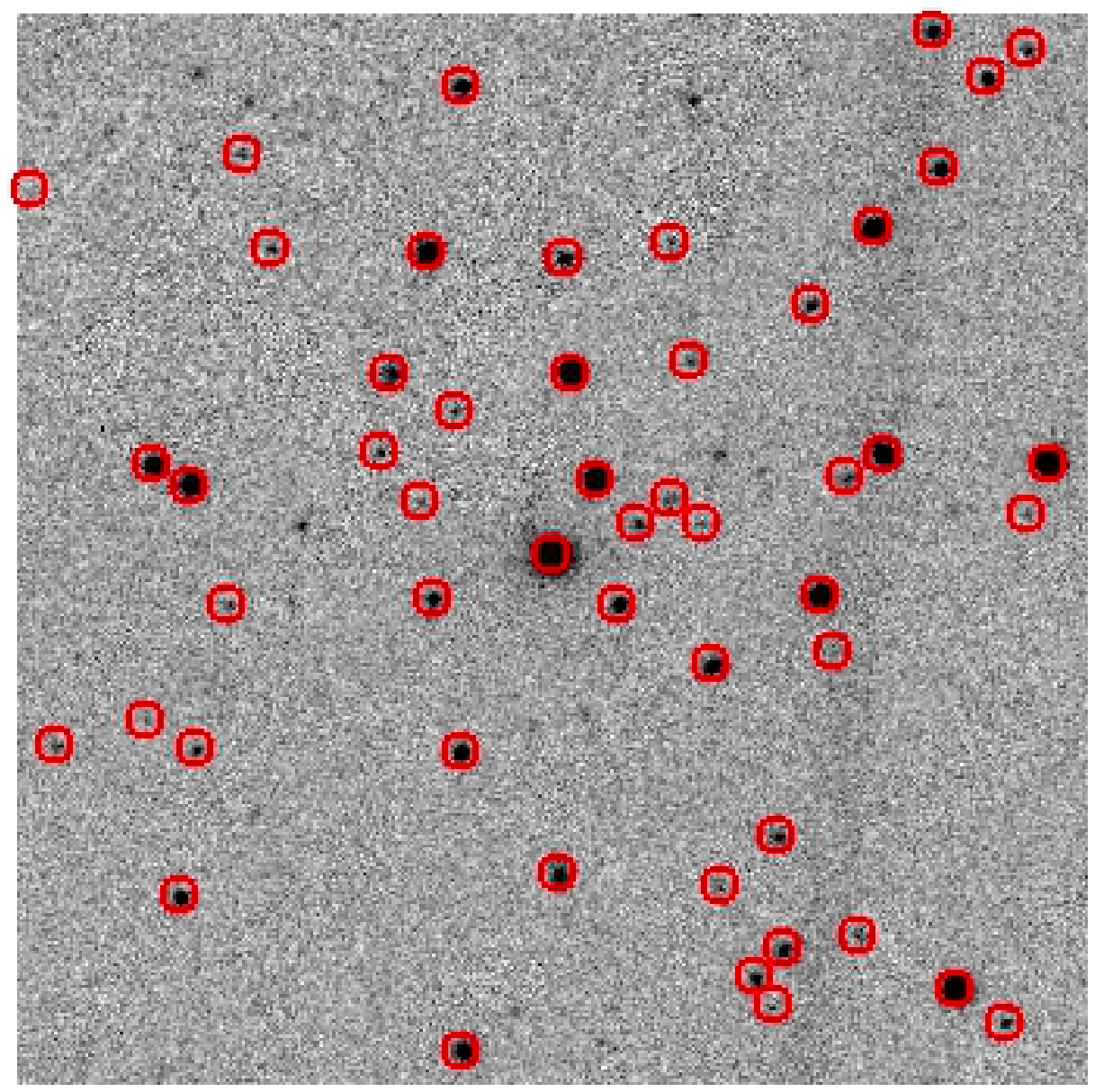} & \includegraphics[scale=0.4,angle=0]{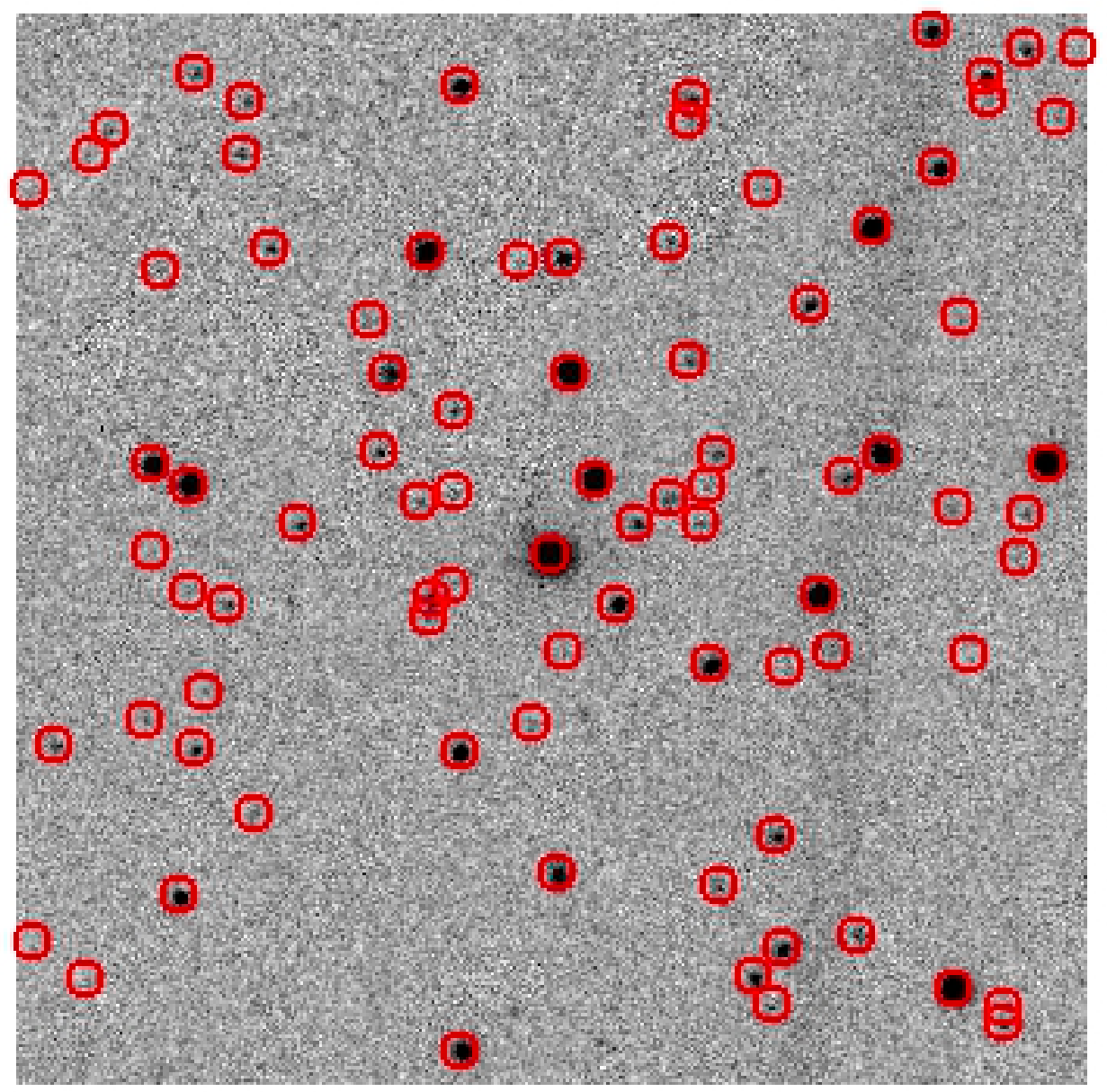} \\
\large{VHS $J$ S/N$>$3 (111 sources)} & \large{VHS $J$ S/N$>$2 (147 sources)} &   \large{VHS $J$ S/N$>$1 (194 sources)}\\
\includegraphics[scale=0.4,angle=0]{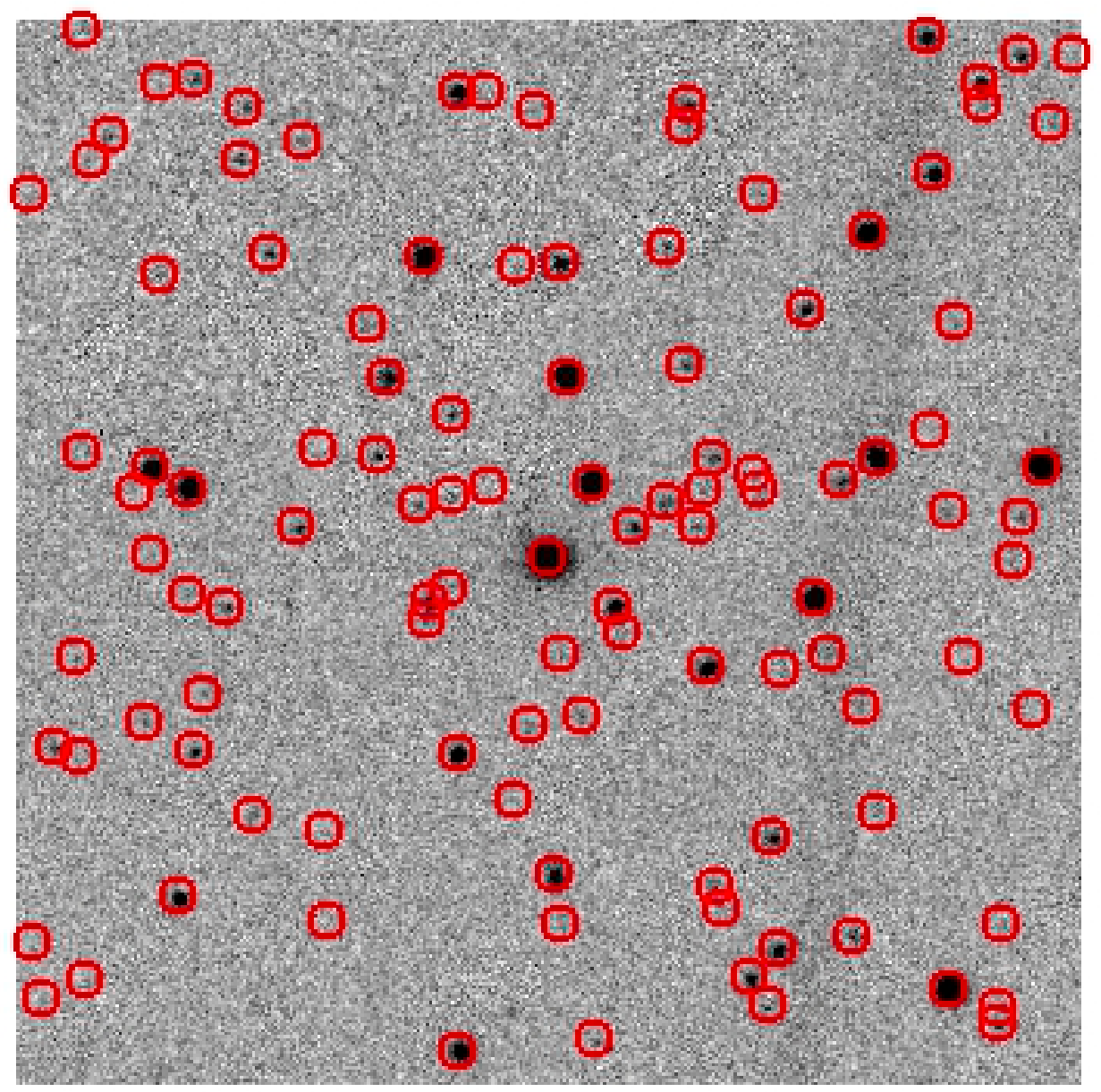} & \includegraphics[scale=0.4,angle=0]{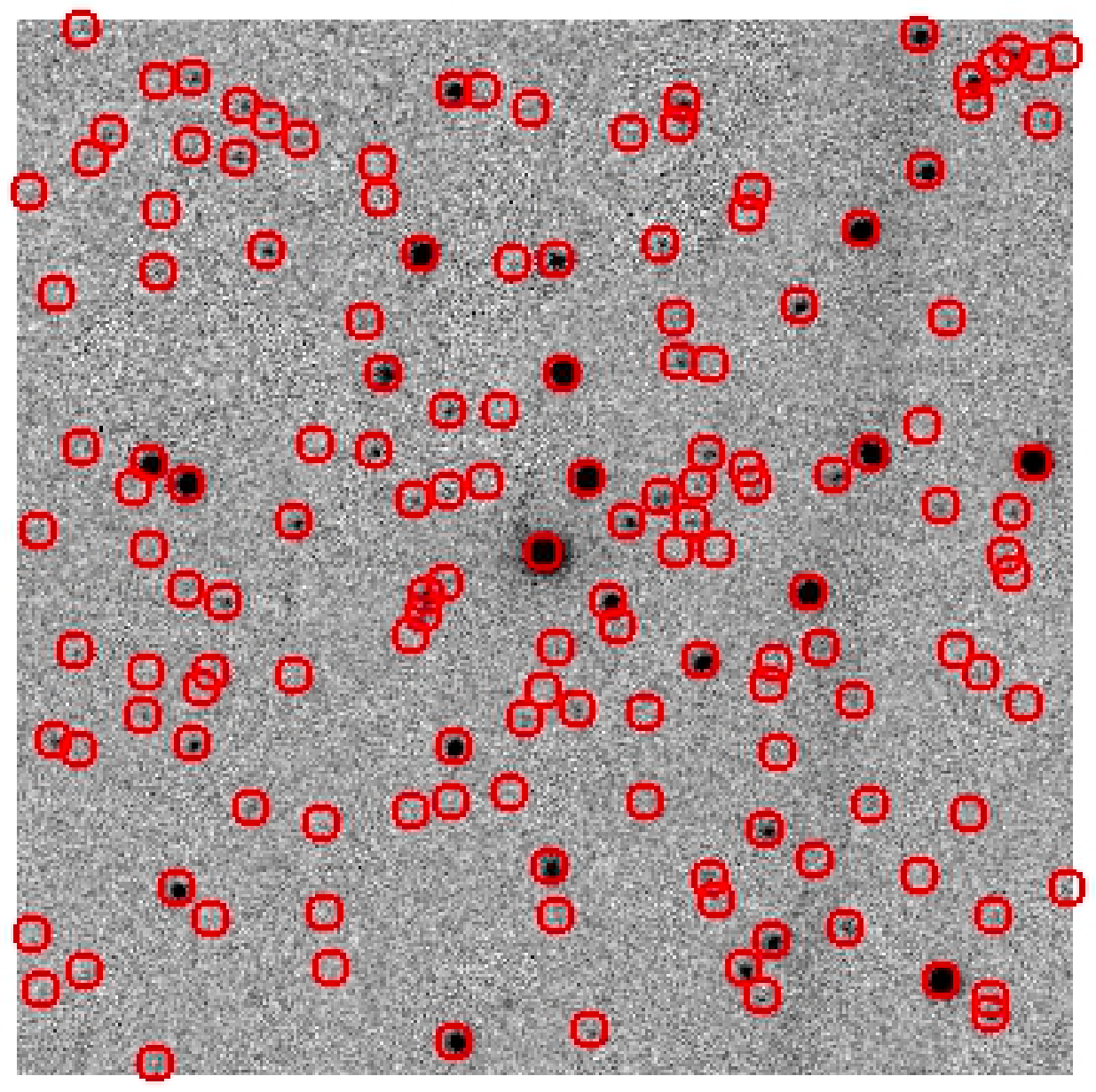} & \includegraphics[scale=0.4,angle=0]{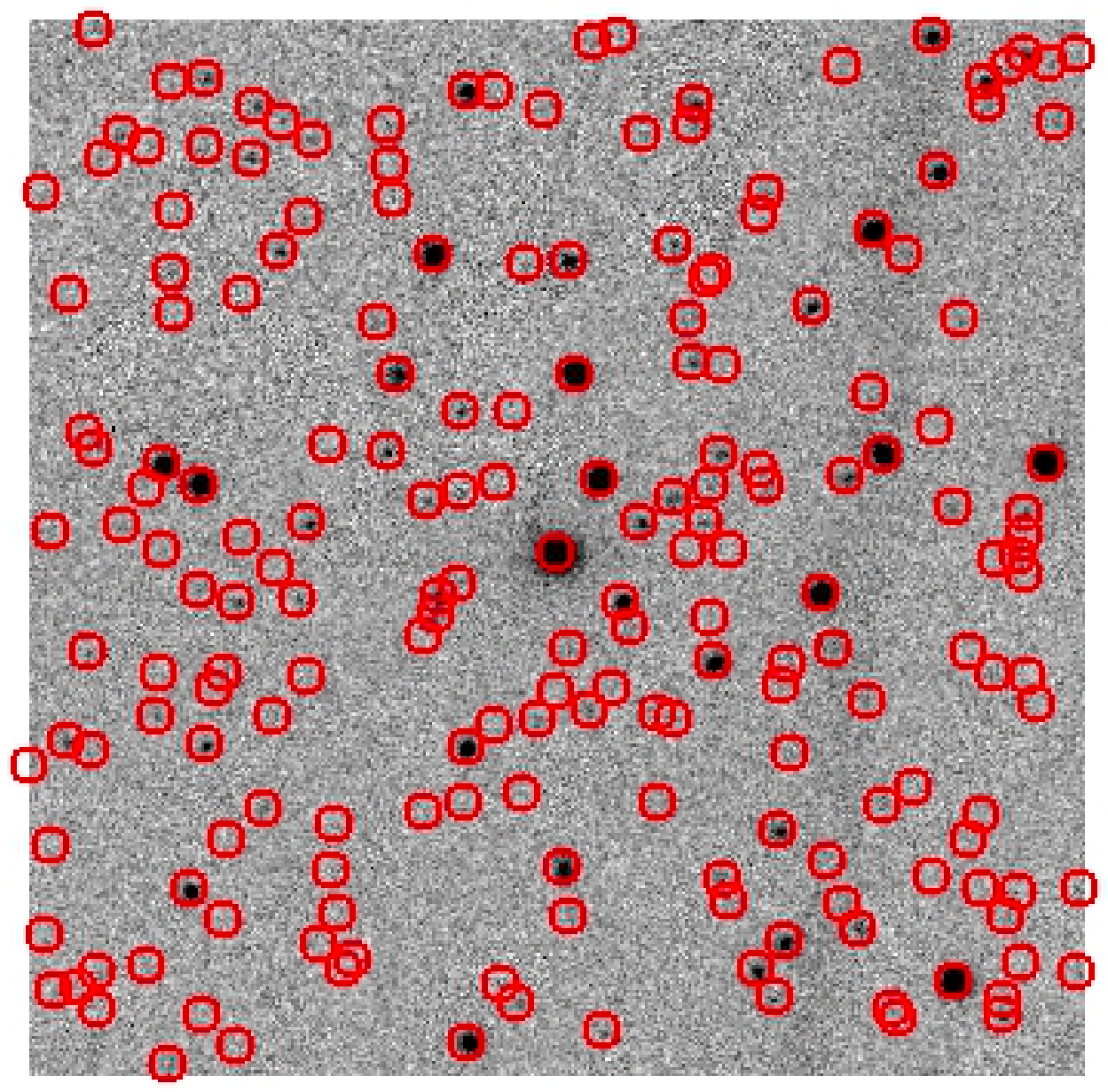}\\
\end{tabular}
\caption{DES $r$-band (top left) and VHS $J$-band images 3$\times$3$^{\prime}$ in size centred on a cluster in the DES SN-X3 field demonstrating the difference in depth between the two surveys. All the DES detections are shown as the blue circles in the DES $r$-band image while each of the VHS images show the sources in our joint DES+VHS catalogue as a function of the $J$-band signal-to-noise ratio. These cutouts demonstrate the significant amount of information present in the VHS images below the 5$\sigma$ threshold used in the construction of VHS catalogues.} 
\label{fig:images}
\end{center}
\end{figure*}

VHS is by design much shallower than DES but nevertheless represents the deepest near infrared survey that will overlap with most of the DES area on completion of DES. Most of the DES galaxies do not have an NIR counterpart in the VHS 5$\sigma$ catalogues. Furthermore, the VHS catalogue production pipeline is different from that used by DES which means that different prescriptions for profile fitting as well as different aperture radii have been used to measure magnitudes in the two surveys. Combining the DES and VHS data in an optimal way requires joint photometry to be carried out on the pixel level data from both surveys. The joint photometry method can make use of the deeper DES detections to extract fluxes from objects that lie below the signal-to-noise threshold used in the construction of the VHS catalogues. The concept is illustrated in Figure \ref{fig:images} where we present DES $r$-band and VHS $J$-band images centred on a galaxy cluster in the SN-X3 field which clearly demonstrates the difference in depth between the two surveys. As is apparent from this figure, there is a significant amount of information in the VHS images below the 5$\sigma$ threshold used to construct the VHS catalogues. Our aim in this paper is to present the first attempt at extracting this extra information directly from the VHS images by making use of the DES detections.

The NIR data represents extra information about the source spectral energy distribution at long wavelengths and, as we will demonstrate later, even in the presence of noise, it can serve as a useful discriminant in breaking degeneracies between galaxy properties. A consequence of performing joint photometry is also so that we ensure that magnitudes are measured consistently across both surveys thus avoiding biases in the optical-NIR colours due to systematic offsets between different magnitude measures for different galaxy populations. 

We now describe the various stages for combining the DES and VHS data. The first step is to select all VHS images overlapping the DECam field-of-view which will then be combined to produce a near infrared image of the same size as the optical DES image. There are six separate VHS tiles that overlap the 3 sq-deg DECam image. Only those images taken under photometric conditions are used corresponding to coadd tiles where the seeing and zeropoint variation between the individual pawprints are smaller than 0.2$^{\prime \prime}$ and 0.2 mag respectively. This ensures that any coadd images where the PSF varies significantly from one pawprint to another, have been removed from the analysis. We note that this seeing and zeropoint cut does not remove any VHS tiles overlapping this single DECam field, but would remove $\sim$17\% of problematic VHS tiles that have been observed and overlap the final DES footprint shown in Figure \ref{fig:area}. Before coaddition, all VHS images overlapping the DECam image need to be scaled to a common zeropoint which is selected to be 25.0. The zeropoints are first adjusted for the effects of different airmass, extinction and exposure time in each image as shown in Eq. \ref{eq:zp} before applying this scaling.

\begin{equation}
\rm{ZP}_{\rm{eff}}=\rm{MAGZPT}-(\rm{airmass}-1.0)\times \rm{ext}+2.5\rm{log}_{10}(\rm{Exp Time})
\label{eq:zp}
\end{equation}

\noindent Using SWarP, the VHS images are then resampled to produce a 30,000$\times$30,000 pixel image with a pixel scale of 0.267$^{\prime \prime}$ per pixel, which therefore directly corresponds to the size and scale of the DECam image. A $LANCZOS3$ resampling algorithm is used. The original VHS images have a pixel scale of 0.341$^{\prime \prime}$ per pixel. The images are also coadded in regions where they overlap using a median coaddition. We note here that resampling the VHS images onto a finer pixel scale introduces correlated noise between the pixels which is not dealt with when magnitude errors are produced by our source extraction software. We come back to this point later in the paper.

Once a VHS coadded image has been produced with the same center, size and pixel scale as the DES image, we run SExtractor in dual-photometry mode using the DES $r$-band image as the detection image and the VHS coadd tile as the measurement image in order to produce a VHS catalogue in the $J$ and $K_s$-bands in the SN-X3 field. The SExtractor parameters used are summarised in Table \ref{tab:sextractor} and mostly correspond to those used during the SV period for the production of DES catalogues. 

\begin{table}
\caption{SExtractor parameters used to produce DES+VHS joint photometry catalogues.}
\begin{center}
\begin{tabular}{cc}
Parameter & Value \\
\hline \hline
DETECT\_MINAREA & 6 \\
DETECT\_THRESH & 1.5 \\
ANALYSIS\_THRESH & 1.5 \\ 
DEBLEND\_NTHRSH & 32 \\
DEBLEND\_MINCONT & 0.005 \\
BACK\_SIZE & 256 \\
BACK\_FILTERSIZE & 3 \\
BACKPHOTO\_TYPE & GLOBAL \\
\hline
\end{tabular}
\end{center}
\label{tab:sextractor}
\end{table}

We would like to check that the photometric properties of our data do not significantly depend on the exact choice of parameters summarised in Table \ref{tab:sextractor}, in particular the exact prescriptions used for background subtraction. In order to do this, we construct histograms of the pixel values of the VHS images both before and after background subtraction. After excluding the high-$\sigma$ tail of this distribution, which corresponds to real sources, we fit a Gaussian to the remaining pixels in order to determine both the shift in the mean pixel value resulting from the background subtraction, as well as the intrinsic width of the pixel distribution. We find that background subtraction results in a shift to the mean pixel value of $0.0143$. By comparison, the best-fit $\sigma$ value of the Gaussian distribution of pixel values is 0.672. In other words, the background subtraction results in a systematic shift in the pixel values that is only $\sim$2\% of the random noise in the images. We have also investigated changes to several other parameters listed in Table \ref{tab:sextractor} - e.g. use of local versus global background subtraction, changes to the background estimation (BACK\_SIZE BACK\_FILTERSIZE) and deblending (DEBLEND\_NTHRSH and DEBLEND\_MINCONT) parameters  - and found no discernible change to the resulting photometry from these changes. Similarly, we investigated differences in the photometry resulting from median versus weighted coaddition of images and once again found that there is a negligible impact on the photometry. These investigations demonstrate that the errors on the photometry are dominated by the random sky noise rather than systematic effects resulting from different processing prescriptions.

We also note that we do not currently include model-fitting to calculate galaxy magnitudes as this involves PSF homogenisation of the VISTA tiles. As the VISTA tiles are constructed from six exposures across 16 sparsely distributed detectors, at any point in the image, there may be contributions from up to 16$\times$6 PSFs. The effects of variable seeing are corrected for when producing the VHS VDFS 5$\sigma$ catalogues using a process known as \textit{grouting}\footnote{http://casu.ast.cam.ac.uk/surveys-projects/software-release/grouting}. However, these corrections are not applied to the VHS tile images that are used in this work. The PSF variation across a coadd tile is not easily modelled using simple smoothly varying functions and the PSF determination is also complicated by the image resampling. PSF and model-fitting to the VHS images are therefore left for future work. We note however that at high redshifts of $z \gtrsim 0.7$ where the VHS data is likely to add the most value to DES, most galaxies are unresolved in the VHS images, which have a typical seeing value of $\sim$1$^{\prime \prime}$. Model-fitting is therefore not appropriate in these cases and simple aperture magnitudes are expected to give reasonable galaxy colours and can be used for a range of science applications as we demonstrate in the following analysis. Throughout this paper, we use the SExtractor $\rm{MAG\_AUTO}$ values for magnitude and colour measurements. 

\section{ASTROMETRY}

\label{sec:astrom}

\begin{figure*}
\begin{center}
\includegraphics[scale=0.5,angle=0]{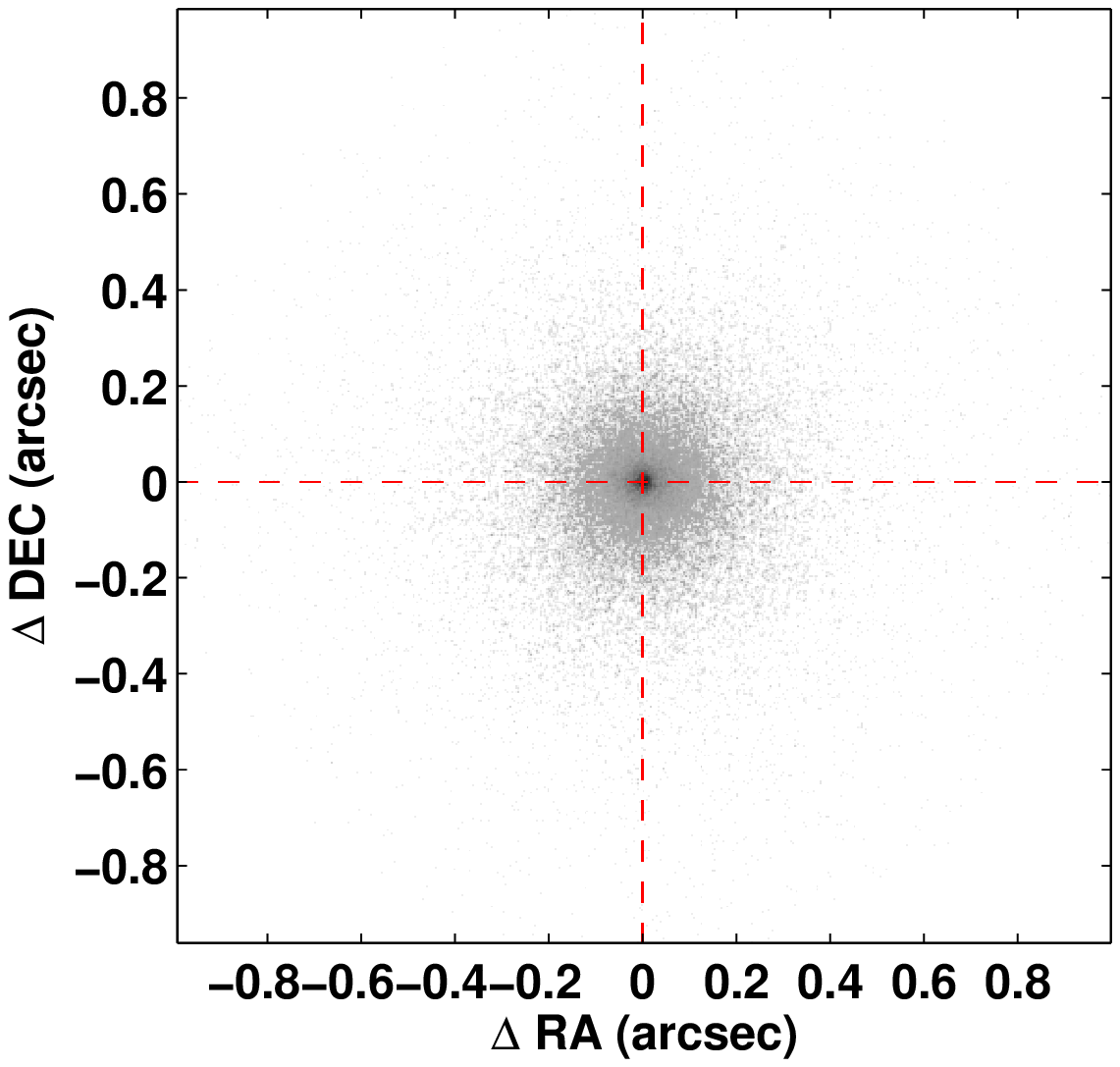}
\includegraphics[scale=0.5,angle=0]{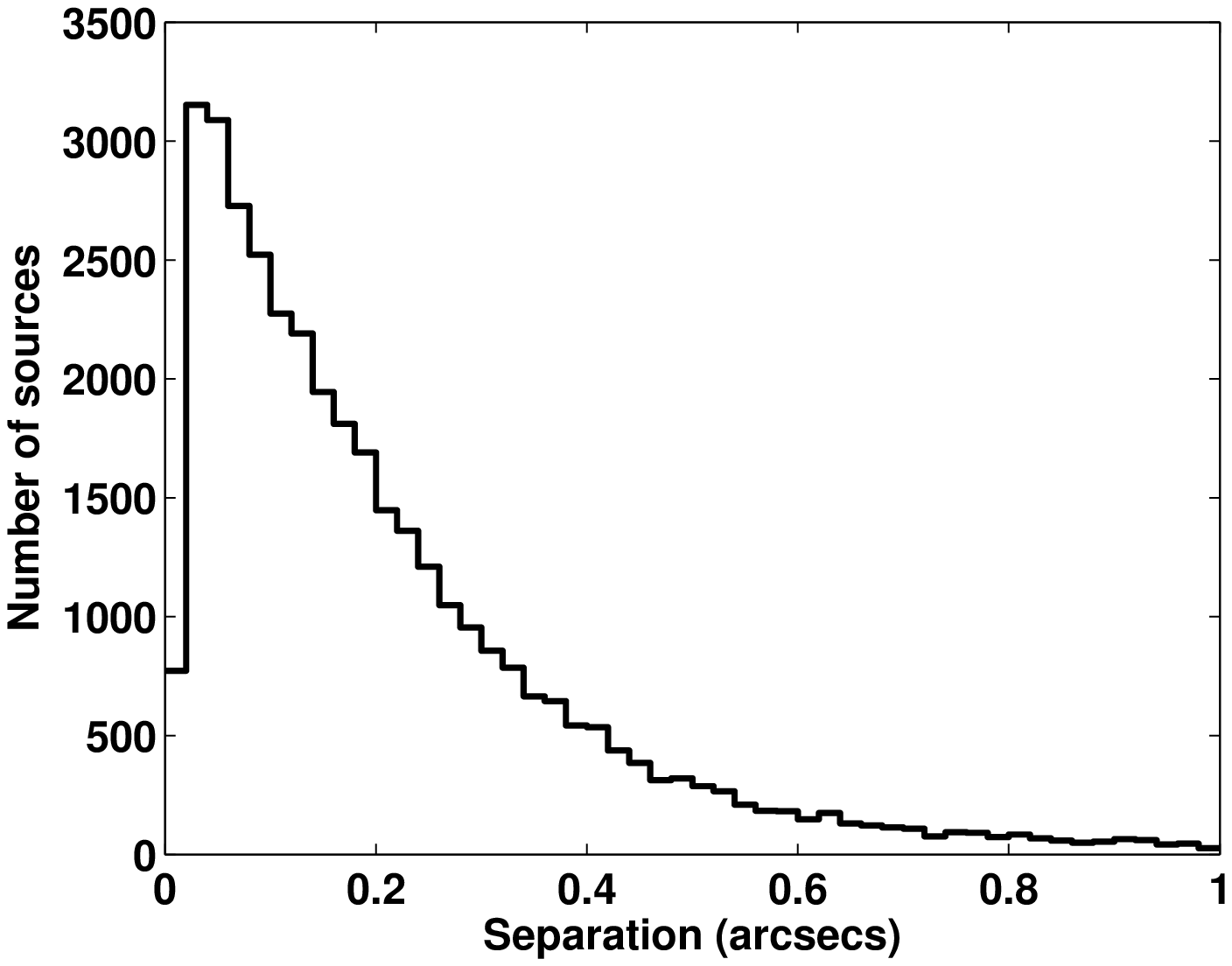}
\caption{\textit{Left:} The 2D distribution of the difference in RA and difference in DEC between DES and VHS. The distribution is strongly peaked close to zero and confirms that the astrometry between the two surveys agrees over the entire DECam field of view. \textit{Right:} The distribution of separations between DES and VHS sources has a mean value of $\sim$0.15$^{\prime \prime}$.}
\label{fig:astrom}
\end{center}
\end{figure*}

Before running the dual photometry on the DES and VHS data, we begin by checking the astrometric calibration between the two surveys. In order to do this, we match the DES catalogues in the SN-X3 field to the VHS 5$\sigma$ catalogues using a matching radius of 1.0$^{\prime \prime}$. The astrometry for VHS has been calibrated using 2MASS while the DES astrometry used in this work has been calibrated using the SCAMP software and the positions from the Sloan Digital Sky Survey Data Release 8. There are $\sim$38,000 sources that are matched between the two catalogues. In Figure \ref{fig:astrom} we plot the 2-dimensional distribution of the difference in the RA and DEC values between DES and VHS. As can be seen, this distribution is strongly peaked at an offset close to zero. The median offsets are: $\delta$RA=RA(DES)-RA(VHS)=0.005$^{\prime \prime}$ and $\delta$DEC=DEC(DES)-DEC(VHS)=-0.002$^{\prime \prime}$. The median absolute deviation in these offsets is 0.12$^{\prime \prime}$ in both RA and DEC corresponding to a standard deviation of 0.18$^{\prime \prime}$. 

Figure \ref{fig:astrom} also shows the histogram of separations for all the objects matched between DES and the VHS catalogues. As can be seen from this figure, the mean separation between DES and VHS sources is 0.15$^{\prime \prime}$. We conclude that the astrometry is therefore consistent across the entire DECam field of view between the two surveys. 

\section{PHOTOMETRY}

\label{sec:photo}

\begin{table}
\caption{Summary of median seeing values in all the DES and VHS filters}
\begin{center}
\begin{tabular}{cc}
Filter & Median Seeing \\
\hline \hline
DES $g$ & 1.24$\arcsec$ \\
DES $r$ & 1.03$\arcsec$ \\
DES $i$ & 0.96$\arcsec$ \\
DES $z$ & 1.12$\arcsec$ \\
DES $Y$ & 1.33$\arcsec$ \\
VHS $J$ & 1.12$\arcsec$ \\
VHS $K_s$ & 1.05$\arcsec$ \\
\hline
\end{tabular}
\end{center}
\label{tab:seeing}
\end{table}

The 3 sq-deg DECam pointing in the SN-X3 field contains $\sim$272,000 sources with at least 6 pixels at 1.5$\sigma$ above the background. The DES $r$-band image is used as the detection image and SExtractor is run in dual photometry mode using this optical image for detection and calculating photometric properties on the near infrared $J$ and $K_s$-band images which directly overlap it. The median seeing values in all the DES bands as well as the two VHS bands, are quoted in Table \ref{tab:seeing} and are around $\sim1\arcsec$. The seeing is slightly worse in the $g$ and $Y$-bands as DES survey strategy dictates that these exposures, which are not being used for weak-lensing science, are taken during the poorer seeing time. The SExtractor parameters for dual photometry are given in Table \ref{tab:sextractor}. The dual photometry therefore allows us to measure VHS $J$ and $K_s$-band fluxes for all sources that are detected in the $r$-band. However, many of these correspond to spurious sources and artefacts in the optical imaging data. In order to remove these, we match the optical sources to the near infrared sources using a matching radius of 1$^{\prime \prime}$ which essentially removes all sources where centroids are poorly determined in the near infrared image. We note that the centroids measured on the NIR images are not actually used for the forced photometry, which uses the positions from the DES detection image. The near infrared $J$-band positions are used for this matching as the $J$-band goes deeper than the $K_s$-band. Sources with poorly determined centroids that are removed, generally correspond to satellite trails, cosmic rays and diffraction spikes around bright stars but also include very faint galaxies where there is no discernible flux in the near infrared and the source detection algorithm can therefore no longer measure a centroid. In crowded cluster fields such as the one shown in Figure \ref{fig:images}, multiple DES sources that are blended into a single VHS source are also removed via this band-merging procedure. The band-merging is therefore used as a simple way of cleaning the catalogues of both spurious detections in the optical and extremely low signal-to-noise sources where even the forced photometry fluxes are likely to be unreliable. 

\begin{figure*}
\begin{center}
\includegraphics[width=8cm, height=6.5cm,angle=0]{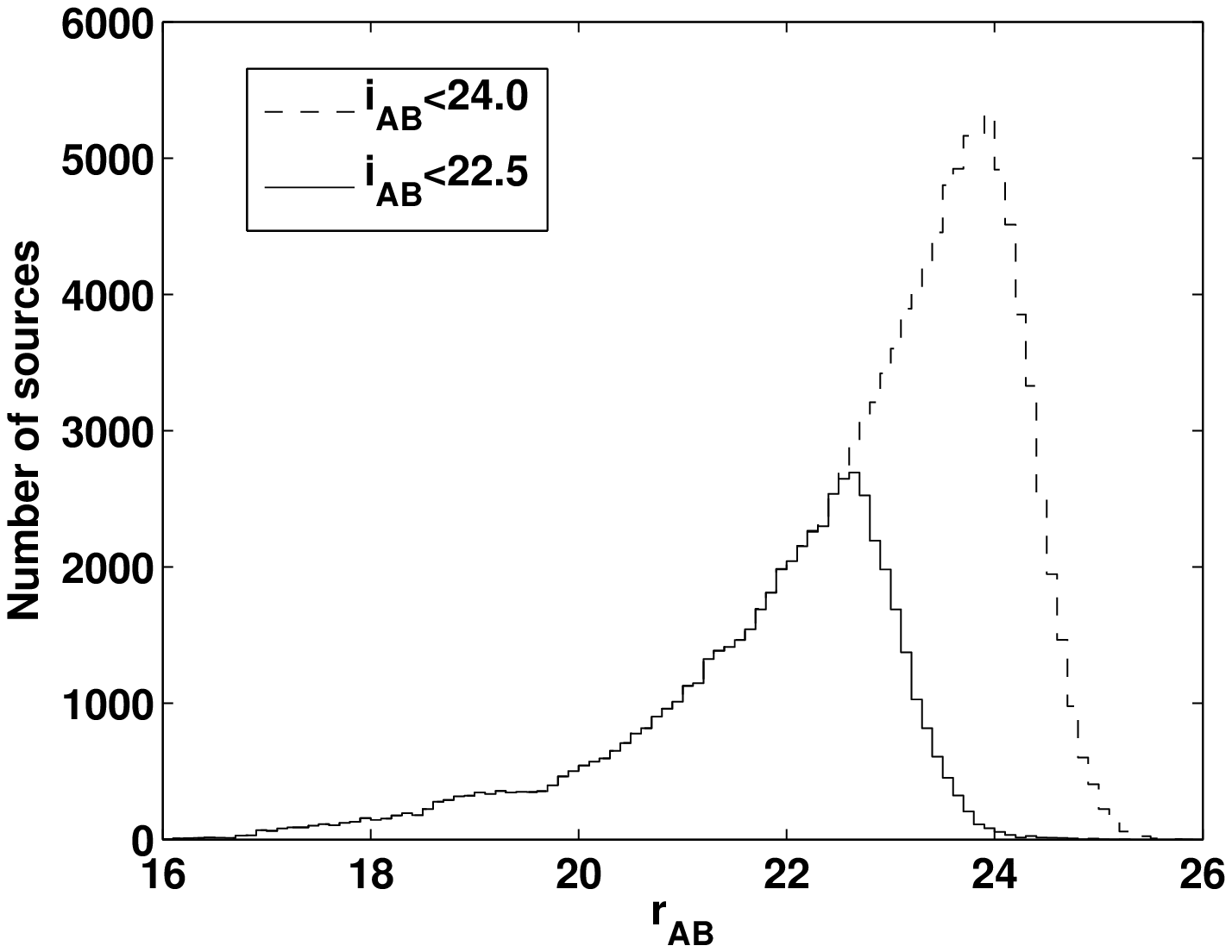}
\includegraphics[width=8cm, height=6.5cm,angle=0]{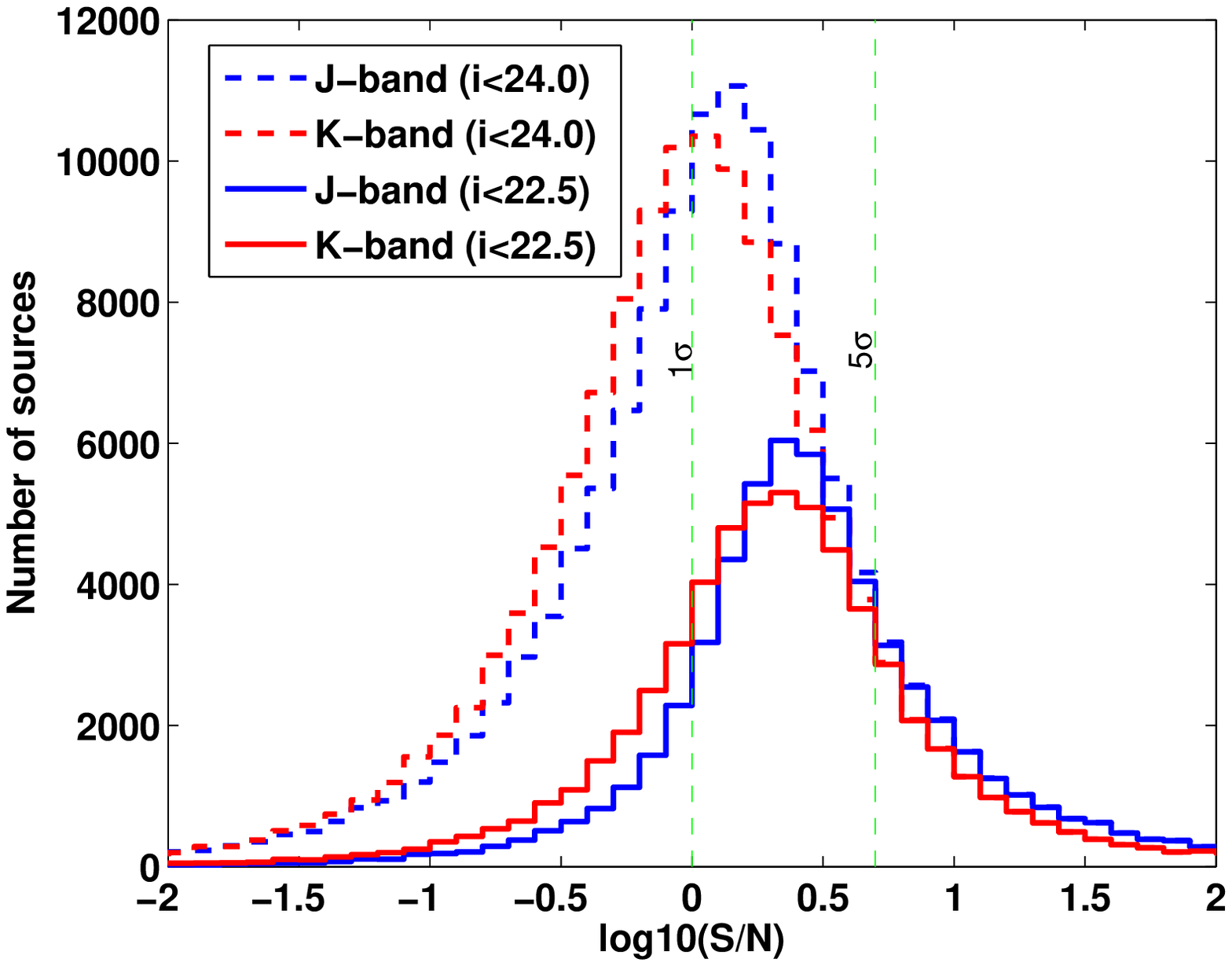}
\caption{\textit{Left:} Histogram of $r$-band magnitudes of DES sources. \textit{Right:} Histogram of signal-to-noise values in the $J$ and $K_s$-bands for sources in our joint DES+VHS catalogue after accounting for correlated noise as described in Section \ref{sec:photo}. The distribution is peaked at a signal-to-noise of $\sim$1. The $J$-band goes slightly deeper and therefore is skewed towards slightly higher signal-to-noise values compared to the $K_s$-band. Selecting sources with $i<22.5$, we find that these distributions peak at a signal-to-noise of $\sim$2-2.5. The VHS catalogues only include objects at signal-to-noise of $>$5 and as can be seen, this corresponds to the brightest tail of DES sources with a median $i$-band magnitude of 20.8. }
\label{fig:snhist}
\end{center}
\end{figure*}

After band merging, the joint DES+VHS catalogue contains $\sim$182,000 sources with VHS photometry - i.e. 67\% of DES sources have measured near infrared fluxes. Out of these $\sim$182,000 sources, only $\sim$38,000 are in the VHS catalogues. Using the DES detections, we have therefore significantly increased the number of sources with near infrared fluxes, albeit now including VHS sources with larger errors on their photometry. Figure \ref{fig:snhist} shows the histogram of both $r$-band magnitudes and signal-to-noise values in the $J$ and $K_s$-band for our final catalogue of $\sim$182,000 sources where the signal-to-noise estimates account for correlated noise in the resampled VHS images as detailed later in this Section. As can be seen the signal-to-noise distribution peaks at $\sim$1 in both bands. Imposing a bright cut of $i<22.5$, we see that the median signal-to-noise is now $\sim$2-2.5. By contrast, the median $i$-band magnitude of those sources present in the 5$\sigma$ VHS catalogues, is $i\sim20.8$. 

\begin{figure*}
\begin{center}
\includegraphics[scale=0.5,angle=0]{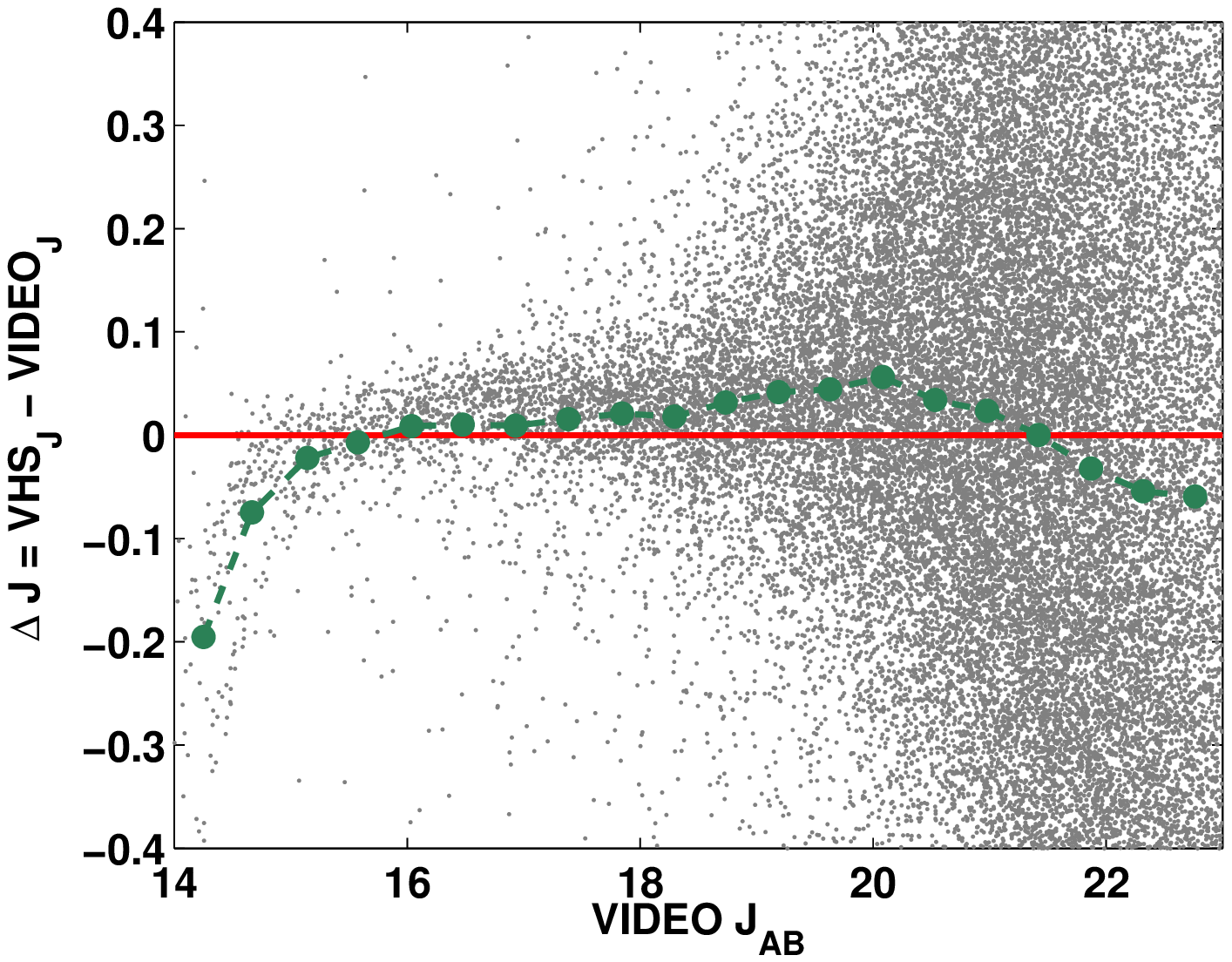}
\includegraphics[scale=0.5,angle=0]{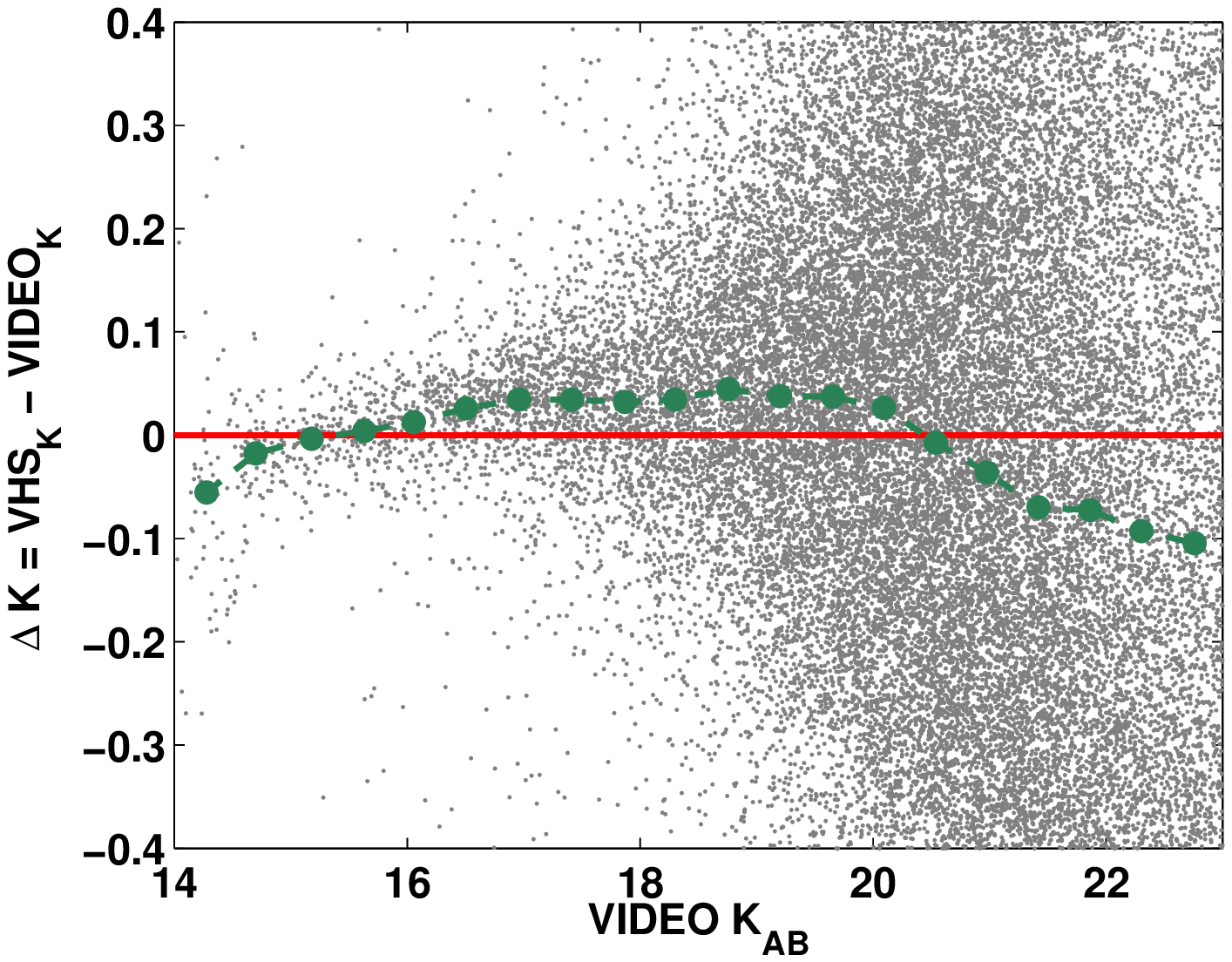}
\caption{Difference between the SExtractor VHS forced photometry magnitudes and deeper VISTA VIDEO catalogue magnitudes in the J and $K_s$ bands as a function of the VIDEO near infrared magnitude. The dashed line corresponds to the median trend whereas the solid line marks an offset of 0 mag. The median offsets are +0.04 mag in $J$ and +0.03 mag in $K_s$ relative to VIDEO at the bright end where the logarithmic magnitude distributions are expected to be roughly Gaussian.}
\label{fig:magoff}
\end{center}
\end{figure*}

In this section, we first begin by assessing the quality of the photometry. To do this, we match the joint DES+VHS catalogue of $\sim$182,000 sources to deep near infrared data from the VISTA VIDEO survey which has magnitude limits of $J_{\rm{AB}}$=24.4 and $K_{\rm{AB}}$=23.8 measured in a 2$^{\prime \prime}$ aperture \citep{Jarvis:13}. VIDEO uses the same camera, telescope and filters as VHS and can therefore be used to verify the VHS photometry. The public VIDEO data release in this field overlapping deep optical data from the CFHTLS, only covers the central 1.0 sq-deg of the 3 sq-deg DECam pointing. There are $\sim$60,000 sources in our band-merged DES+VHS catalogue that also have VIDEO photometry. The VIDEO data release includes SExtractor output catalogues with $\rm{MAG\_AUTO}$ measurements, which are compared to the photometry generated in this paper. In Figure \ref{fig:magoff} we plot the difference in magnitude in the $J$ and $K_s$ bands between our VHS magnitudes and the VIDEO magnitudes for these sources. We note that in the low signal-to-noise regime, the errors on the magnitudes are non-Gaussian and hence the median difference between the magnitudes is not expected to be zero. At faint magnitudes, the VHS estimates begin to get systematically brighter relative to VIDEO. At bright magnitudes and high signal-to-noise however where the magnitude errors are approximately Gaussian, the magnitudes between the two surveys are consistent at the level of $\sim$0.04 mag. At very bright magnitudes, the effects of saturation can be seen in the $J$-band. While the detector integration time in the $J$-band is 30s for VIDEO, 15s integration times have been used instead for VHS so it is possible that bright objects with $J<15$ could be saturated in the longer VIDEO $J$-band exposures. 

\begin{figure*}
\begin{center}
\includegraphics[scale=0.5,angle=0]{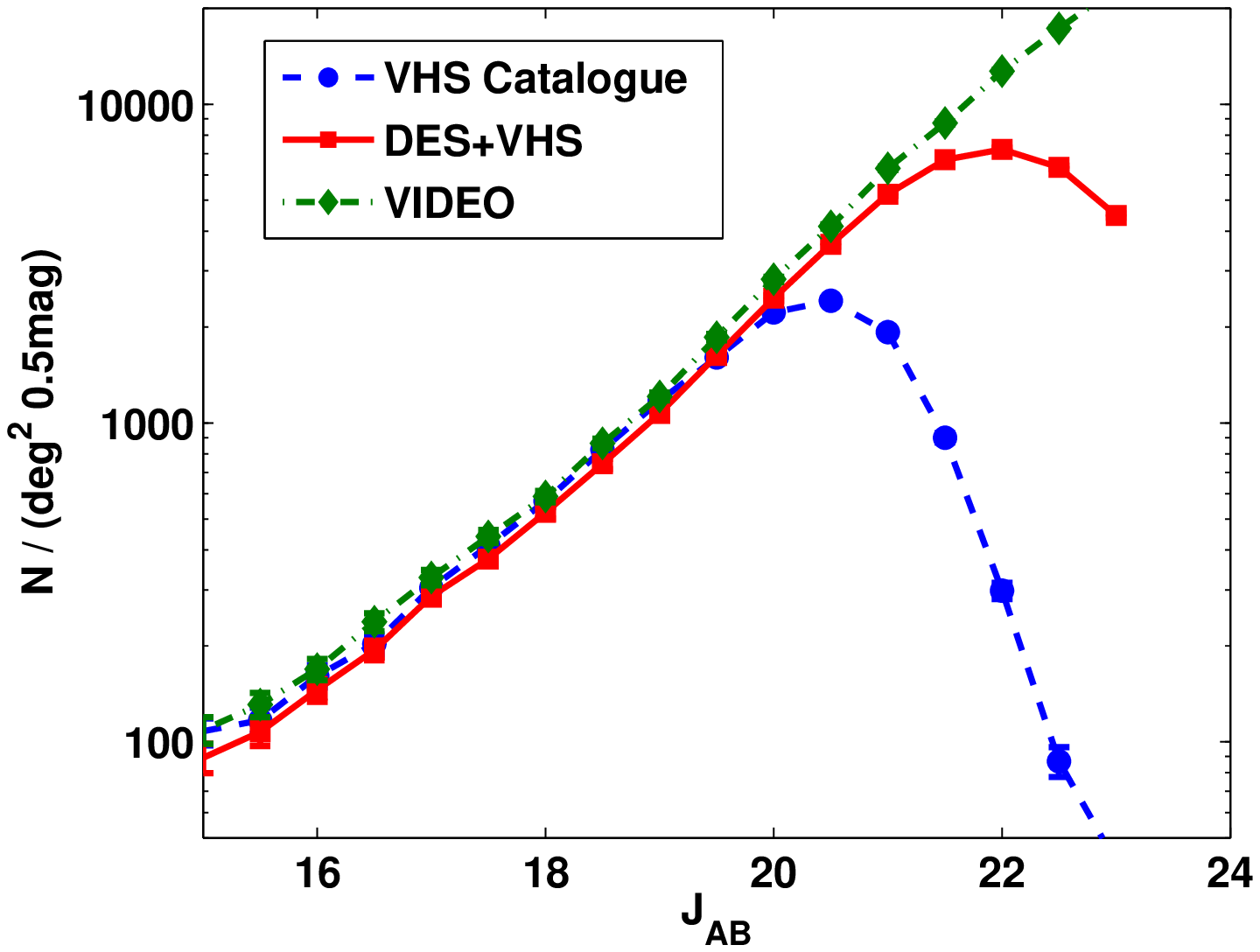}
\includegraphics[scale=0.5,angle=0]{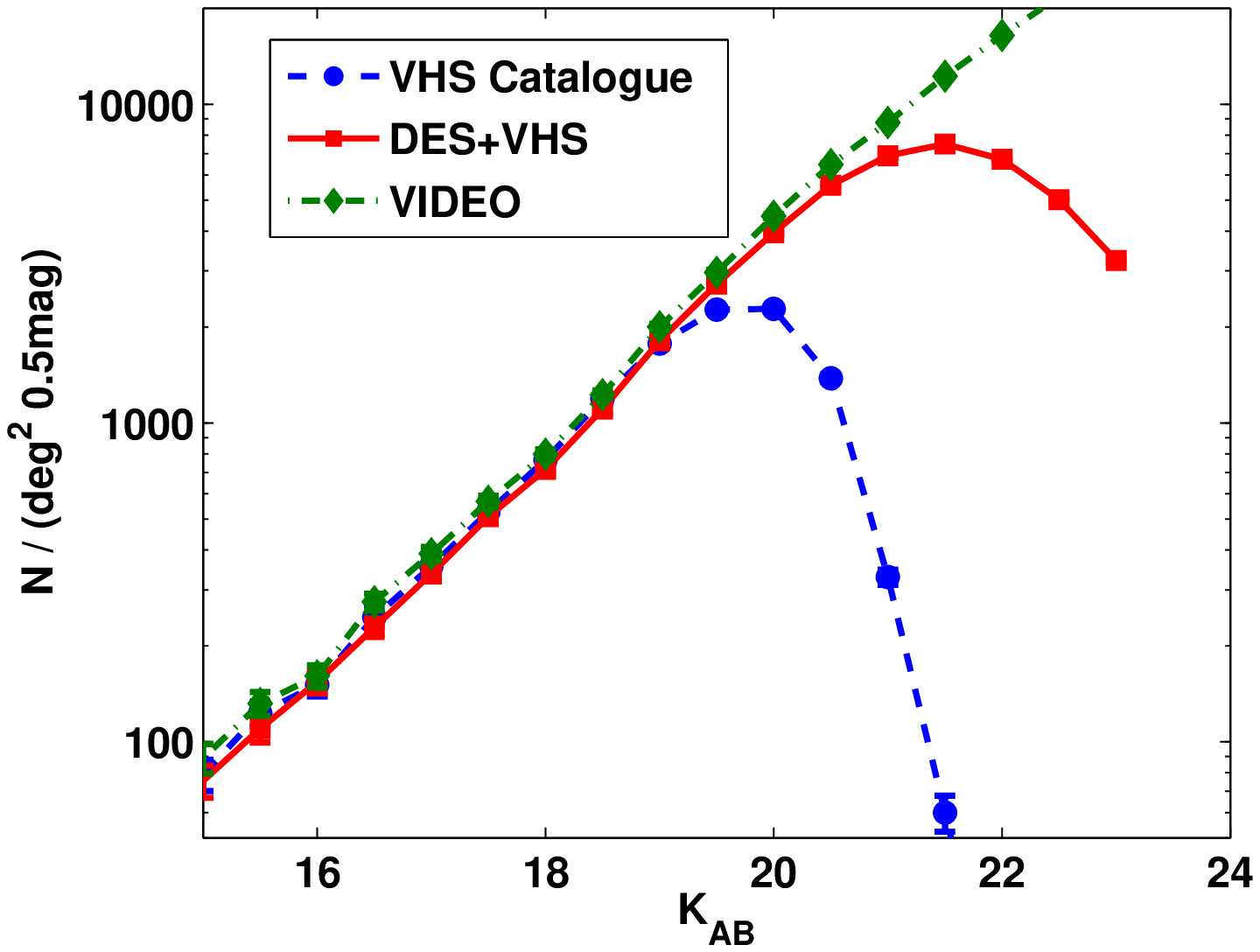}
\caption{VHS $J$ and $K_s$-band number counts produced by running dual photometry on the VHS coadded images using the DES detection images (DES+VHS) compared to both VHS catalogue counts over the same area as well as deeper data from the VIDEO survey. All three distributions show good agreement at the bright end. The number counts are dominated by the extended sources and from this plot, the 80\% completeness for the VHS catalogues is $J=19.9$ and $K=19.3$ (appropriate for galaxies). The corresponding numbers for the DES+VHS dual photometry catalogue presented in this paper are $J=21.2$ and $K=20.9$.}
\label{fig:ncounts}
\end{center}
\end{figure*}

In Figure \ref{fig:ncounts} we plot the number of sources as a function of $J$ and $K_s$-band magnitude in the VHS catalogues, VIDEO catalogues and the DES+VHS joint forced photometry catalogues produced in this work. The number counts of all three catalogues agree very well at the bright-end. The small offset of $\sim$0.04 mag between VHS and VIDEO noted earlier, does not affect Figure \ref{fig:ncounts} where the numbers are calculated in bins of 0.5 mag. The VHS catalogue source counts begin to turn over at reasonably bright magnitudes. Using the VIDEO data, we derive 80\% completeness limits of 19.9 and 19.3 for the $J$-band and $K_s$-band respectively. Note that these are shallower than the median 5$\sigma$ point-source depths of the survey quoted in Section \ref{sec:vhs} as the number counts in Figure \ref{fig:ncounts} are dominated by galaxies and these completeness limits are therefore appropriate for extended sources. Even for point sources, the 5$\sigma$ catalogue flux-limits typically translate to a completeness limit of $\sim$50\% on average. The corresponding 80\% completeness limits for the DES+VHS dual photometry catalogues are $J=21.2$ and $K=20.9$. Using the DES detections, we have therefore been able to extend the near infrared coverage to $\sim$1.5 mag fainter.  

As stated earlier, the resampling of the VHS images onto the finer DECam pixel scale, introduces correlated noise between the pixels which is not accounted for by SExtractor. The SExtractor magnitude errors are therefore under-estimated. In order to calculate the factor by which these errors are under-estimated, we compare the SExtractor magnitude errors to those produced by the VHS pipeline for high signal-to-noise sources. The VHS pipeline correctly accounts for correlated noise in the pixels. We find that typically the SExtractor magnitude errors are under-estimated by a factor of $\sim$3.5-4. This factor is constant as a function of magnitude and just corresponds to the ratio of the pixel-to-pixel RMS in the original VHS images versus the resampled VHS images used for the dual photometry. As a conservative estimate, we therefore scale all the VHS photometric errors output by SExtractor by a factor of 4, in order to correctly account for correlated noise between the pixels of the resampled VHS images. Throughout the paper, the noise estimates used and quoted therefore correctly account for the correlated noise. 

\begin{figure*}
\begin{center}
\includegraphics[scale=0.5,angle=0]{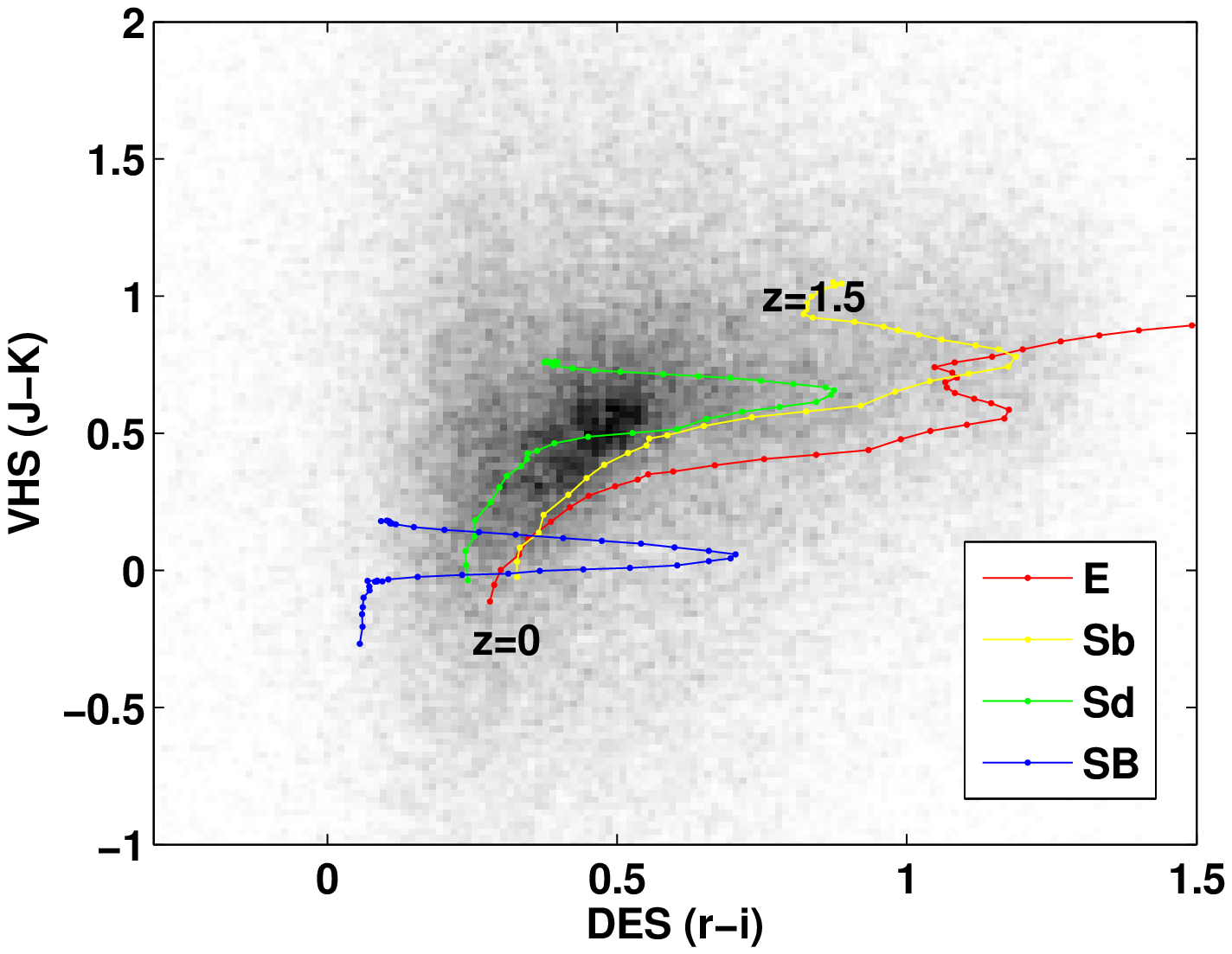}
\includegraphics[scale=0.5,angle=0]{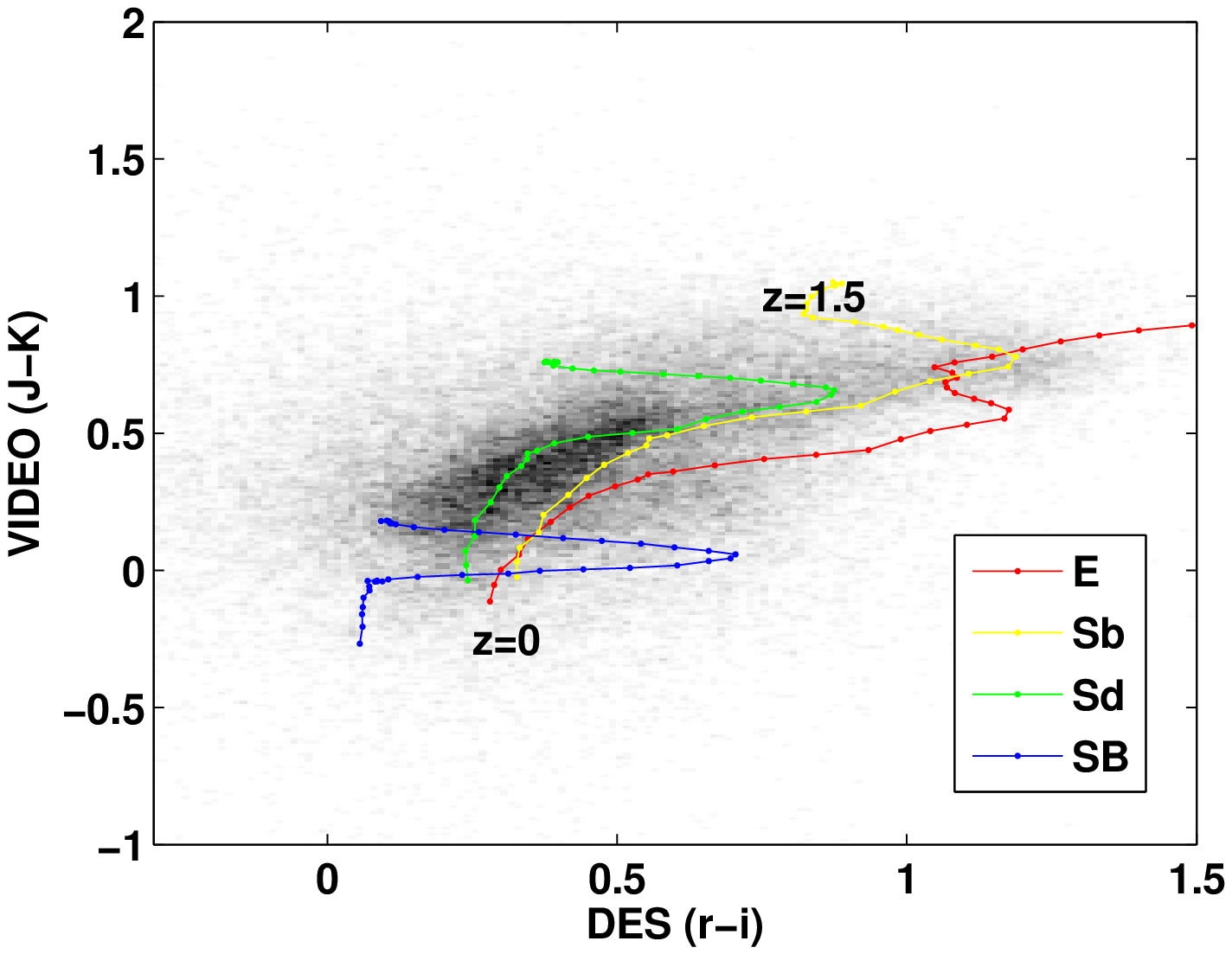}
\caption{The grey clouds show source densities of galaxies in the DES+VISTA $(r-i)$ versus $(J-K)$ colour-colour plane for both our DES+VHS forced photometry catalogue (left) and DES+VIDEO catalogue (right). The coloured tracks mark the expected colours of galaxies from \citet{Jouvel:09} for four different galaxy templates spanning the full range of galaxy types in those catalogues. The tracks go from redshift 0 to 1.5 with points marking redshift intervals of 0.04. The observed colours of our galaxies in the DES+VISTA catalogue, overlap well with the predicted colours of galaxies from \citet{Jouvel:09} but the larger spread in the $(J-K)$ colours in VHS relative to VIDEO results from the larger photometric scatter on the shallower VHS data.}
\label{fig:colour}
\end{center}
\end{figure*}

We now consider the colours of galaxies in our joint DES+VHS catalogues. We select galaxies using a crude star/galaxy separation cut of $i\_spreadmodel > 0.0028$. The use of the new $spread\_model$ parameter \citep{Desai:12} in performing star-galaxy separation for DES will be discussed in more detail in Section \ref{sec:sg}. While this cut does not necessarily represent the most optimal separation criterion, we show in Section \ref{sec:sg} that it allows reasonable classifications that are sufficient for our aims and this cut is therefore used throughout this paper. There are $\sim$125,000 galaxies over the 3 sq-deg pointing when using this selection. The $(r-i)$ versus $(J-K)$ colours of these galaxies is shown in Figure \ref{fig:colour}. Here we also plot the DES+VISTA colours of galaxies for four different galaxy templates from \citet{Jouvel:09} spanning the entire range in galaxy types in these mock catalogues. The equivalent colour-colour distribution for the deeper VIDEO survey has also been plotted. As can be seen from the figure, the main locus of data points in this colour-colour plane overlaps the galaxy tracks thus demonstrating that the forced photometry pipeline is producing reasonable colours for galaxies. Although the deeper VIDEO data have less scatter in the $(J-K)$ colour by a factor of $\sim$3 compared to VHS, Figure \ref{fig:colour} clearly demonstrates that the peak of the galaxy distribution in both datasets roughly coincides, and correspond to a $z\sim0.5$ star-forming galaxy in terms of its optical+NIR colour. 

\section{STAR-GALAXY SEPARATION}

\label{sec:sg}

For the purposes of this paper, we would like to be able to separate stars and galaxies in the DES and VHS catalogues. Within DESDM, star-galaxy separation makes use of the new $spread\_model$ parameter within SExtractor \citep{Desai:12, Soumagnac:13}. Point sources which are unresolved in the DES images should have a $spread\_model$ value close to zero whereas more extended sources will have larger $spread\_model$ values. Several improvements to the classification are currently under investigation (e.g. \citealt{Soumagnac:13}) and are not the subject of the current study. In this work we choose to perform star/galaxy separation using a $spread\_model$ threshold of 0.0028. 

\begin{figure*}
\begin{center}
\includegraphics[scale=0.35,angle=0]{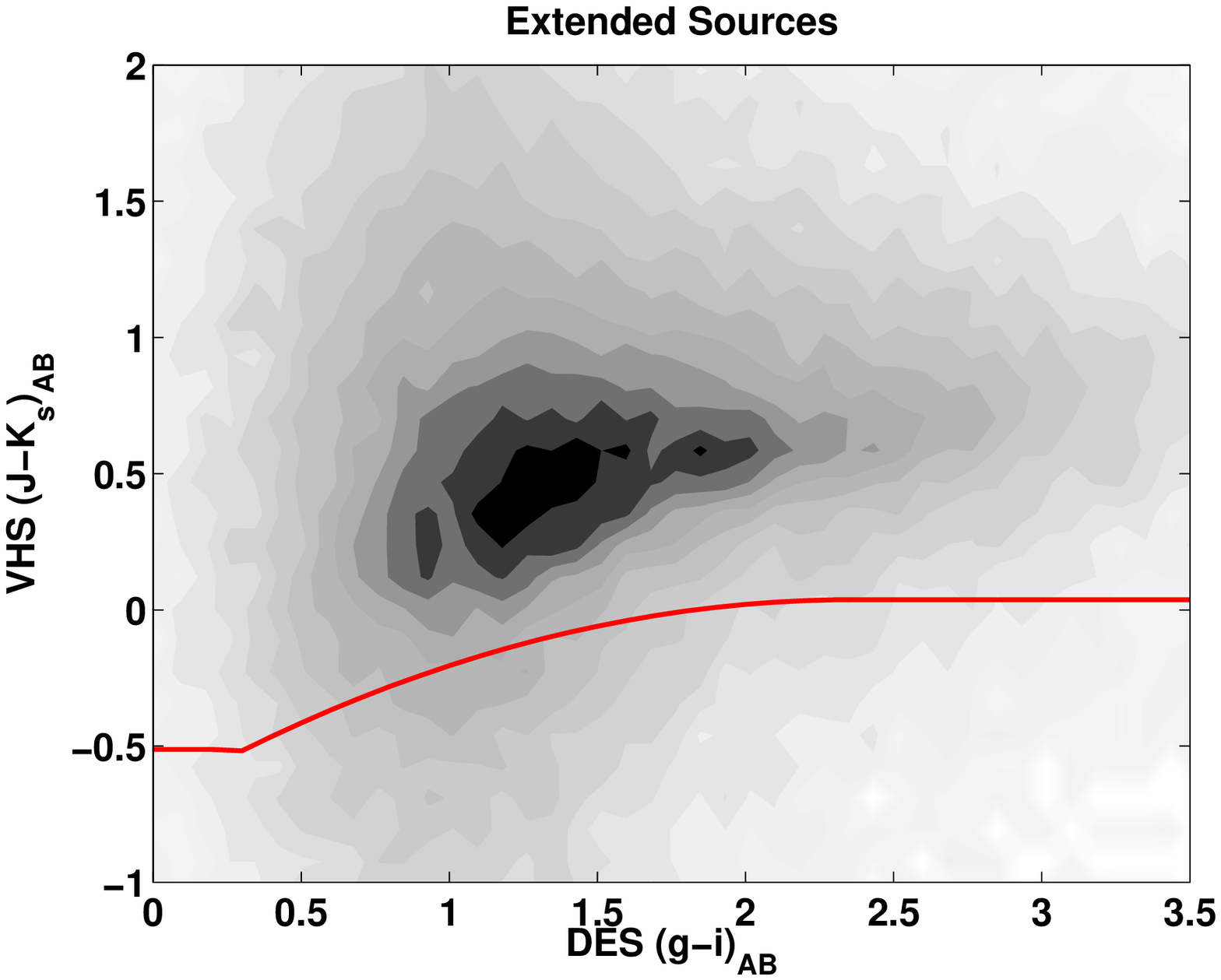}
\includegraphics[scale=0.35,angle=0]{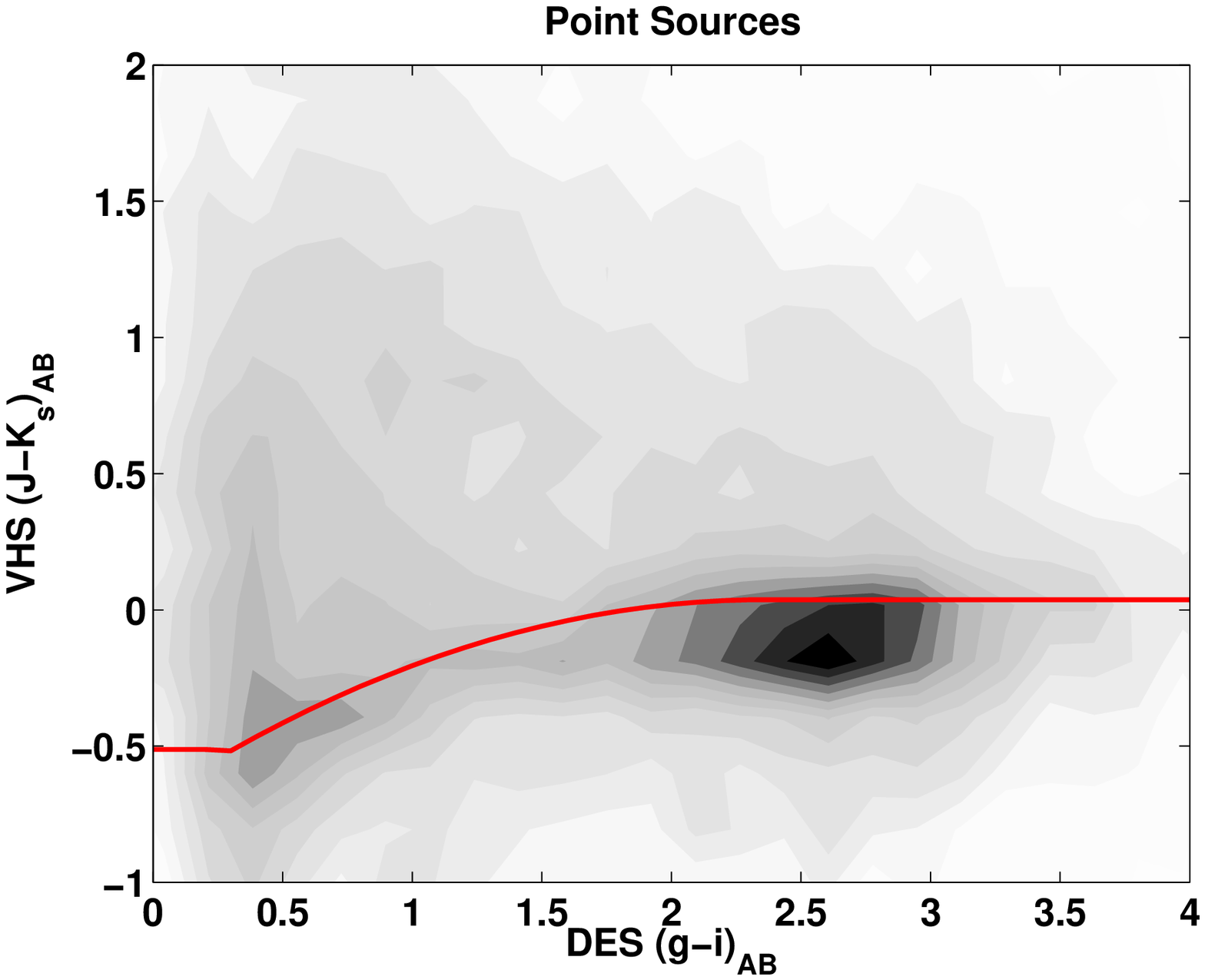}
\caption{Source density plot in the $(g-i)$ versus $(J-K)$ colours of both extended sources (assumed to have $i\_spread\_$model$>$0.0028; left) and point sources (assumed to have $-0.003<i\_spread\_model<$0.0028; right) in the DES+VHS catalogues produced in the SN-X3 field. The locus used to separate the two populations by \citet{Jarvis:13} is marked in each panel by the thick line. While the extended sources generally lie above this locus, the point sources lie below as expected.}
\label{fig:sg}
\end{center}
\end{figure*}

We wish to test the effectiveness of the above $spread\_model$ cut in separating stars from galaxies in the DES+VHS catalogues. The combination of optical and near infrared photometry can effectively be used for this as stars and galaxies have been shown to separate well in the $(g-i)$ versus $(J-K)$ colour-colour plane \citep{Baldry:10}. We split point sources and extended sources in our catalogue using the criteria: $-0.003<i\_spread\_model<0.0028$ and $i\_spread\_model>0.0028$ respectively. This results in $\sim$40,000 stellar sources and $\sim$125,000 galaxies, and the two populations are plotted in the $giJK$ colour-colour plane in Figure \ref{fig:sg}. The figure also shows the stellar locus from \citet{Jarvis:13} which is essentially the \citet{Baldry:10} locus used to separate the two populations, with appropriate small offsets to convert from the UKIDSS $K$-band to the VISTA $K_s$-band. As can be seen from the figure, the majority of the point sources do indeed fall below the line while the extended sources lie above the line. The significant population of point sources with blue $(g-i)$ but red $(J-K)$ colours and that lie above the locus, are mainly expected to be quasars. Using this colour locus for star/galaxy separation, we find that 85\% of extended sources in DES with $i\_spread\_model>0.0028$, lie in the correct portion of the $giJK$ colour-colour plane. 

Optimum star-galaxy separation criteria for DES sources will be investigated in forthcoming studies and the above criteria are not intended to give highly pure or highly complete samples of either population, which are necessary for example for much of the cosmological analysis intended with DES. Nevertheless, we can conclude based on the optical+NIR colours that a simple $i\_spread\_model>0.0028$ cut is sufficient for the the aims of this paper, which are to assess the potential of the NIR data from VHS in enhancing DES science.  

\section{DES+VHS PHOTOMETRIC REDSHIFTS}

\label{sec:photoz}

\begin{figure*}
\begin{center}
\includegraphics[scale=0.4,angle=0]{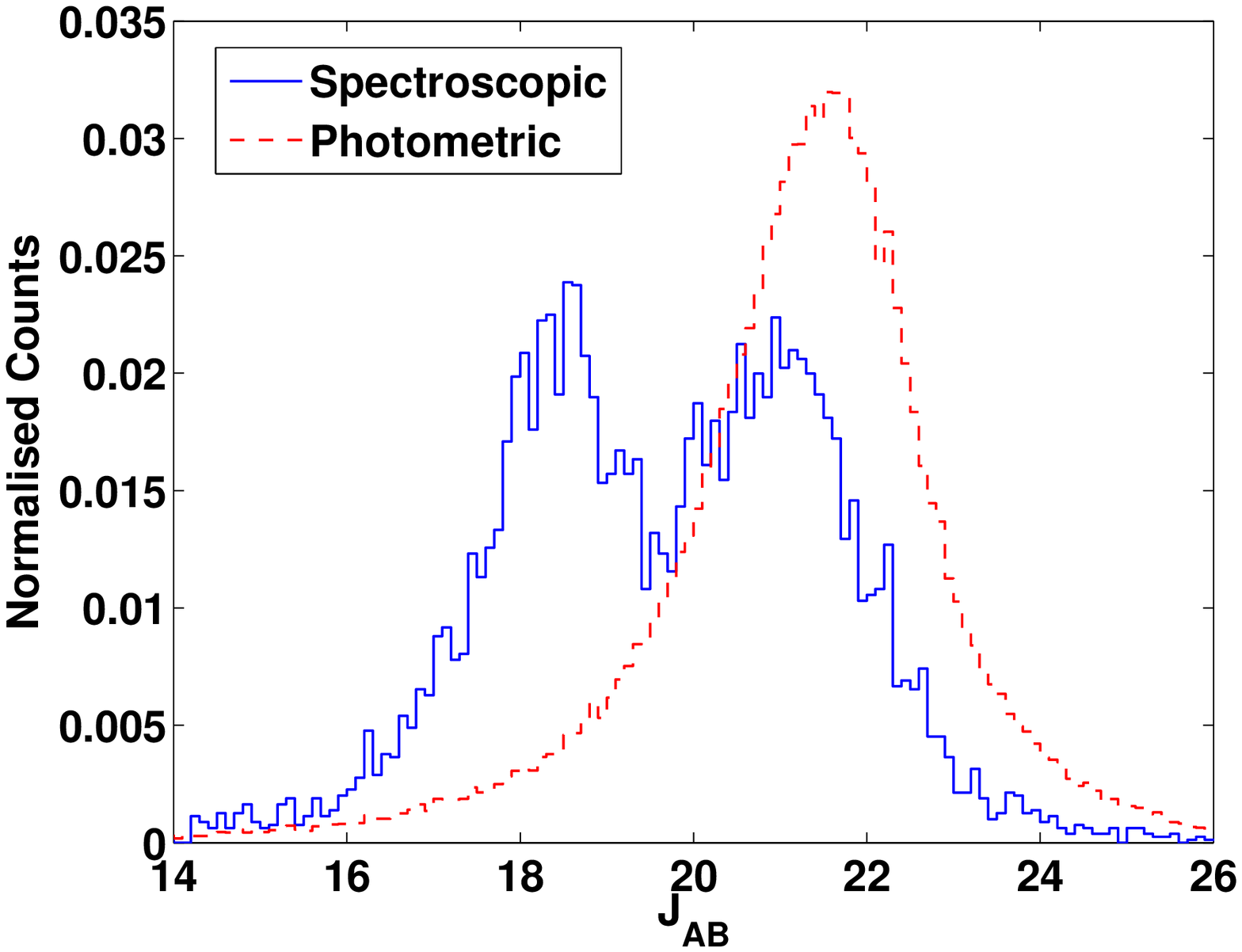}
\includegraphics[scale=0.4,angle=0]{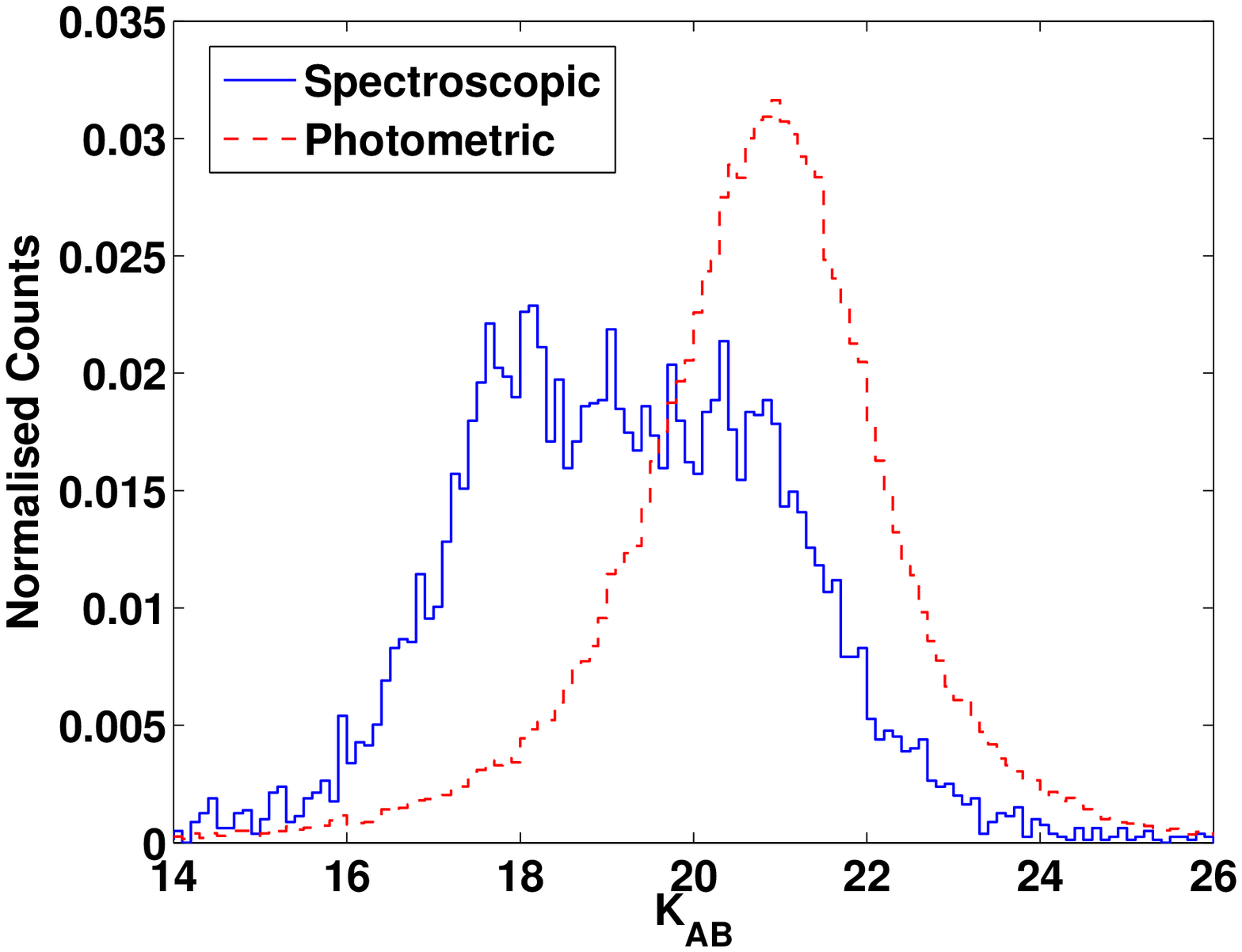}
\caption{Magnitude distributions (normalised by the total number of galaxies) for both the spectroscopic dataset used for photometric redshift analysis and the full photometric dataset produced by our forced photometry pipeline. The spectroscopic dataset is clearly biased towards brighter galaxies with evidence for a bimodal distribution due to the superposition of galaxies from bright surveys such as SDSS, 6dF and GAMA and fainter surveys such as VVDS-Deep and VIPERS, in our master redshift catalogue.}
\label{fig:JKhist_spec}
\end{center}
\end{figure*}

In \citet{Banerji:08} it was demonstrated using early DES+VHS simulations that the addition of near infrared photometry from VHS to the DES optical data can result in a substantial improvement in the photometric redshift scatter of DES galaxies at high redshifts. In particular, these early simulations indicated that the improvement should be $\sim$30\% at $z\gtrsim1$ and this improvement in photometric redshift performance could also result in a significant improvement in large-scale-structure measurements with DES. In this Section, we would therefore like to test how the DES photometric redshifts change on addition of the NIR VHS data. Spectroscopic redshifts are available in the SN-X3 field from OzDES as well as several other spectroscopic surveys as described in Section \ref{sec:ozdes}. There are $\sim$9800 galaxies overlapping our DES+VHS catalogues in this field, all with high confidence spectroscopic redshift measurements. Out of these, $\sim$6800 are in the VHS $>$5$\sigma$ catalogues so the bright galaxies are clearly over-represented in this spectroscopic sample relative to the full photometric dataset. More than 70\% of these galaxies are also at $z<0.7$ which is not representative of the final redshift distribution of the DES photometric sample. Figure \ref{fig:JKhist_spec} plots the magnitude distributions in the $J$ and $K_s$-bands for both the spectroscopic and photometric datasets and clearly illustrates this point. The photometric redshifts presented in this section should therefore be interpreted with caution as, for the single field used in this pilot study, they depend strongly on both the spectroscopic sample being used as well as the photometric redshift algorithm. 

For the purposes of this paper, we run a single template fitting code: LePhare \citep{Arnouts:05} on these galaxies in order to assess the photometric redshift performance. LePhare has been commonly used for the analysis of photometric redshifts in several deep field galaxy surveys such as VIDEO \citep{Jarvis:13} and COSMOS \citep{Ilbert:09} and use of the same algorithm allows us to benchmark our results against these deeper datasets over smaller areas of sky. Furthermore, the spectroscopic samples in the single DECam pointing used in this pilot study are currently too small to provide robust and independent training, validation and testing sets for empirical photometric redshift estimators. A thorough comparison of different photometric redshift algorithms has been carried out by \citet{Sanchez:14} for the DES Science Verification data and we note here that the results presented in this Section do not necessarily constitute the best photometric redshifts that can currently be obtained for DES. Training-set based estimators produce more accurate photometric redshifts than template-fitting methods like LePhare when large training samples overlapping the photometric sample exist. Rather, the aim here is to look at the relative difference in photometric redshift performance for DES and DES+VHS galaxies in this single field.  

The templates used for our photometric redshifts have been optimised
for the COSMOS photo-$z$ library as described in \citet{Ilbert:09}. This library is composed
of 31 templates ranging from ellipticals to starbursts. It includes three elliptical galaxy templates
and seven spiral galaxy templates generated by \citet{Polletta07}.
In addition, there are 12 starburst templates from \citet{BC03} with ages between 0.03
and 3 Gyr. Additional dust-reddened templates are also constructed for the intermediate and late-type populations 
using the extinction law from \citet{Calzetti:00} with E(B-V) values in the range 0.1 to 0.3. Emission line fluxes have 
been additionally incorporated \citep{Ilbert:09, Jouvel:09}. The addition of these emission line fluxes significantly
improves the ability to reproduce colours of star-forming galaxies. Figure \ref{fig:colour} shows that these templates overlap the optical+NIR colour-colour plane spanned by our photometric data and are therefore appropriate for the photometric redshift analysis carried out here. Finally, adaptive photometric offsets are also calculated in each of the photometric pass bands using the spectroscopically confirmed galaxies, in order to correct for small systematic offsets in the magnitudes. 

We begin by making use of a set of relatively simple metrics to assess how the VHS photometry impacts the photo-$z$ performance. The aim is to assess the relative difference in photometric redshift performance between the DES only and DES+VHS catalogues. For these purposes, we evaluate the bias on the photometric redshift $<\Delta z/(1+z_s)>$, the scatter on the photometric redshift by looking at the normalised median absolute deviation (NMAD) of the difference between the photometric redshift and spectroscopic redshift:

\begin{equation}
\sigma_z (NMAD)=\rm{median}\left(1.48\times \frac{|z_p - z_s|}{(1+z_s)}\right)
\label{eq:nmad}
\end{equation}

\noindent In addition we also define the outlier fraction to be the fraction of objects where $|\Delta z|/(1+z)>0.15$. These are standard definitions employed in several other studies \citep{Ilbert:09, Jarvis:13}, that allow us to characterise the photometric redshift performance relative to other surveys. With LePhare, we find that the bias decreases from $-0.013$ with the DES-only data to $-0.007$ with DES+VHS. The scatter on the photometric redshift decreases from 0.058 to 0.052 and the fraction of outliers decreasing from 13\% to 10\%. In \citet{Jarvis:13}, photometric redshifts were evaluated for CFHTLS deep optical data and CFHTLS+VIDEO optical+NIR data where the scatter decreased from 0.025 to 0.023, a 10\% improvement consistent with our results. The outlier fraction too dropped from 7\% to 4\% in that study. Thirty band photometric redshifts in the deep COSMOS field reach a scatter of 0.012 for comparison although this scatter increases to 0.06 when NIR data is added \citep{Ilbert:09}. 

The above statistics while useful for comparison to other studies, are not representative of the accuracy with which photometric redshifts will be measurable for all DES+VHS galaxies, given that the spectroscopic sample is highly biased towards bright magnitudes and low redshifts. In order to overcome this bias, we therefore adopt a weighting method whereby each galaxy in the spectroscopic sample is assigned a weight so that the distribution of magnitudes in this weighted spectroscopic sample is comparable to the distribution of magnitudes in the final photometric sample. Weights are computed using the nearest neighbour estimator of \citet{Lima:08} where the spectroscopic galaxy weight is assigned based on the ratio of the local density of the spectroscopic and photometric samples in the multi-dimensional magnitude plane. More details can be found in \citet{Sanchez:14}. For comparison to the results obtained in \citet{Sanchez:14} for a variety of different photometric redshift algorithms, we also now compute the mean bias, $\sigma_{68}$ parameter and fraction of 2 and 3-$\sigma$ outliers as defined in that paper. To assess how accurately the photometric redshift distribution reproduces the spectroscopic redshift distribution, we calculate the N$_{\rm{poisson}}$ and KS-metrics introduced in \citet{Sanchez:14}. Finally, in line with DES science requirements, we remove 10\% of galaxies with the largest photometric redshift errors before these metrics are calculated. With these cuts, our results are analogous to those presented as Test1 of \citet{Sanchez:14} although we note that our spectroscopic dataset is different from the one used in that work. 

We begin by looking at the overall change in the metrics for DES and DES+VHS using LePhare. These are summarised in Table \ref{tab:metrics} and errors on these estimates are calculated using 100 bootstrap samples. We find that most of the metrics show slight improvements with the addition of VHS data with the most significant improvement seen in the scatter parameter, $\sigma_{68}$. The dispersion on each of these metrics is however almost always smaller with the addition of the near infrared data. In other words, quantities like the bias and scatter can be calculated more robustly with the DES+VHS data than with DES alone. We note that the relatively modest change to the photometric redshift performance with the inclusion of the VHS photometry is expected given the properties of the galaxies that make up the majority of our photometric catalogue. In Figure \ref{fig:colour} we have seen that the average DES galaxy is a $z\sim0.5$ star-forming galaxy with blue optical colours. Given the significantly deeper DES data relative to VHS, such a galaxy has much smaller errors on its DES photometry relative to VHS. The template-fitting $\chi^2$ minimisation method therefore assigns a much larger weight to the DES photometry when calculating photometric redshifts. We also calculate the photometric redshift metrics for only those galaxies that are $>$5$\sigma$ detections in the $J$ and $K_s$-bands and present these in Table \ref{tab:metrics}. While larger improvements in the bias and scatter are now seen as expected, the errors on these parameters are also much larger as these parameters are being estimated using a much smaller number of galaxies.

\begin{table*}
\caption{Summary of the different metrics used in \citet{Sanchez:14} to assess the photometric redshift accuracy both for catalogues with and without VHS data. Metrics are computed for the weighted spectroscopic sample with photometric redshifts from LePhare and after removing 10\% of galaxies with the largest photo-$z$ errors. Errors on the metrics are derived using 100 bootstrap samples.}
\begin{center}
\begin{tabular}{lcccccc}
& $\bar{{\Delta z}}$ & $\sigma_{68}$ & $f_{2\sigma}$ & $f_{3\sigma}$ & N$_{\rm{poisson}}$ & KS \\
\hline
DES & $-0.032\pm0.005$ & $0.086\pm0.003$ & $0.055\pm0.004$ & $0.025\pm0.002$ & $9.473\pm0.563$ & $0.147\pm0.009$ \\
DES+VHS & $-0.038\pm0.003$ & $0.076\pm0.002$ & $0.052\pm0.003$ & $0.025\pm0.002$ & $9.126\pm0.585$ & $0.139\pm0.008$ \\
DES ($JK$ SNR$>$5) & $0.038\pm0.007$ & $0.095\pm0.009$ & $0.049\pm0.009$ & $0.019\pm0.004$ & $1.963\pm0.028$ & $0.192\pm0.019$ \\
DES+VHS ($JK$ SNR$>$5) & $0.015\pm0.007$ & $0.068\pm0.008$ & $0.059\pm0.009$ & $0.026\pm0.006$ & $1.371\pm0.032$ & $0.129\pm0.019$ \\
\hline
\end{tabular}
\end{center}
\label{tab:metrics}
\end{table*} 

\begin{figure*}
\begin{center}
\includegraphics[scale=0.4,angle=0]{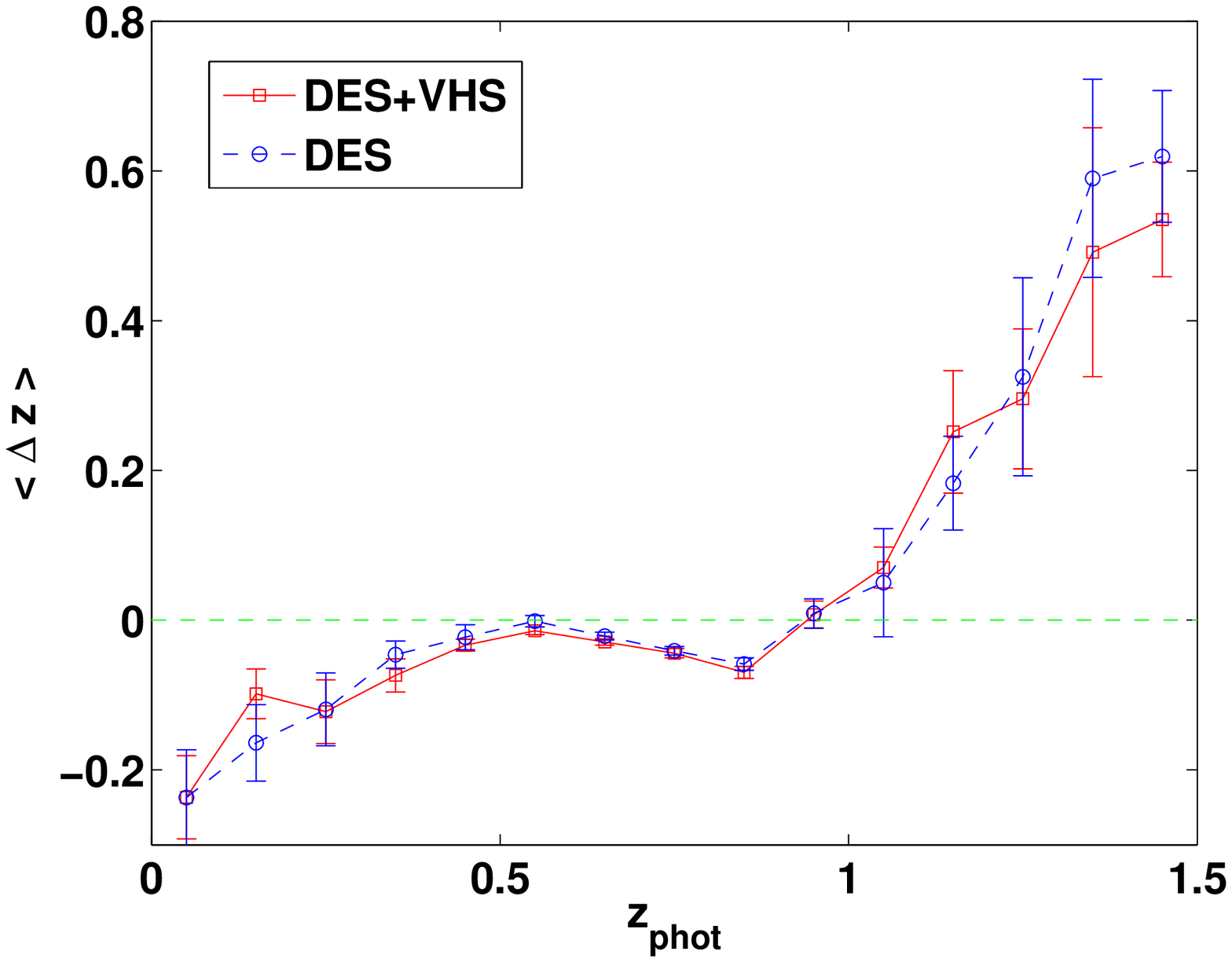}
\includegraphics[scale=0.4,angle=0]{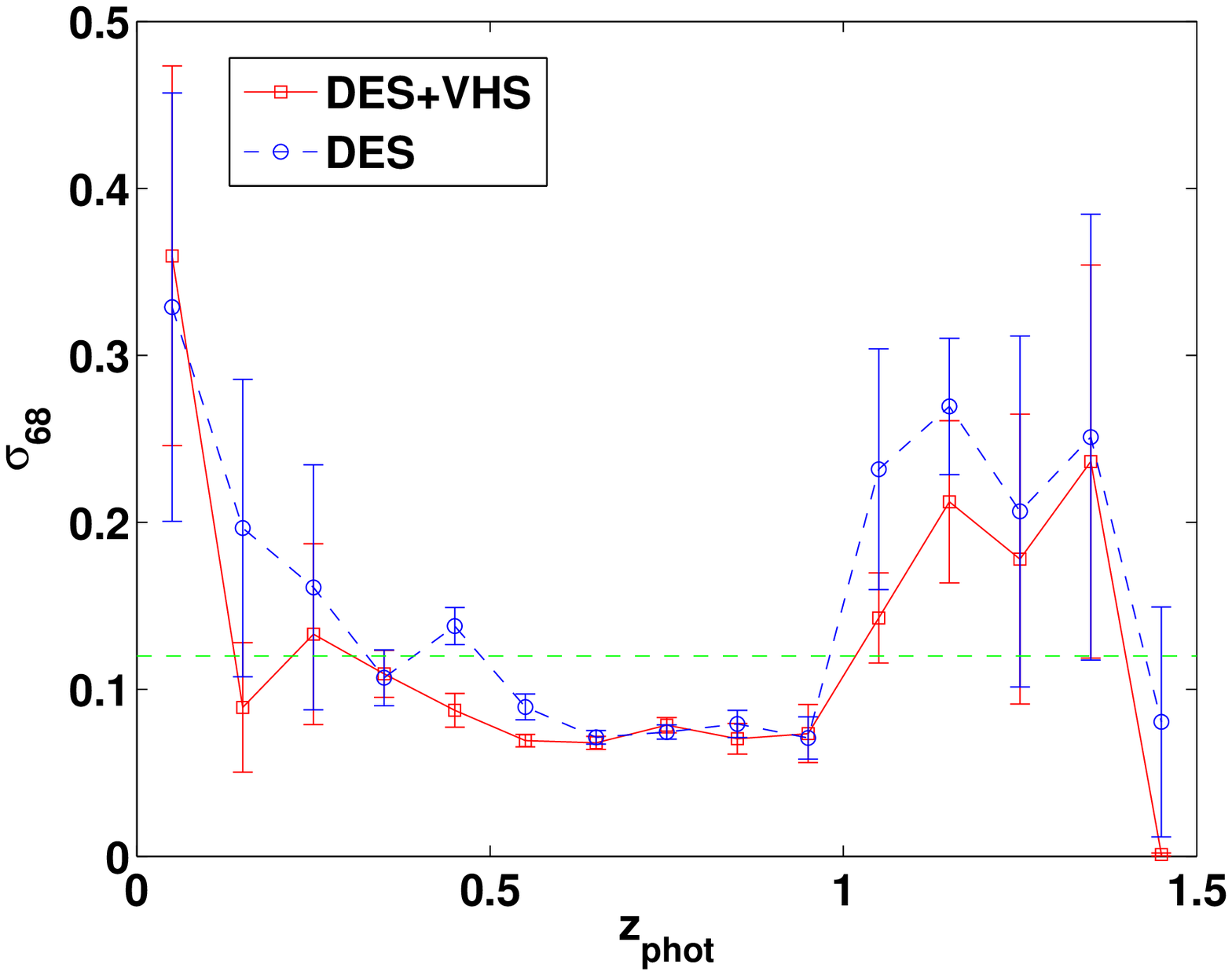}
\caption{The bias on the photometric redshift (left) and the scatter, $\sigma_{68}$ (right) in bins of photometric redshift for both the DES and DES+VHS estimates of the photometric redshift. Galaxies have been weighted as detailed in Section \ref{sec:photoz} before constructing these plots and 10\% of the galaxies with the largest errors on their photometric redshift have been removed. Errors on the points are computed using 100 bootstrap samples.}
\label{fig:metrics}
\end{center}
\end{figure*}

More interesting perhaps than these overall metrics are the trends in these as a function of redshift. Figure \ref{fig:metrics} shows both the bias and $\sigma_{68}$ as a function of photometric redshift both for DES and DES+VHS estimates of the photometric redshift. These plots are produced after weighting the spectroscopic sample of galaxies as discussed earlier and also after removing 10\% of galaxies with the largest photometric redshift errors. The bias looks relatively similar for the two datasets although as noted earlier, this bias can be estimated with a smaller error when the VHS data are included in the photometric redshift estimates. Looking at the scatter we find that, in this single field there is a significant improvement in $\sigma_{68}$ in the redshift bin $z\sim0.4-0.5$, although these results may be prone to cosmic variance. At $z\lesssim0.4$ and $z\gtrsim1$ too there is evidence for a consistent drop in the scatter on addition of VHS data to DES, across multiple redshift bins, although currently the error bars are large due to small number statistics in this single DECam pointing, which does not contain enough spectroscopic galaxies at the edges of the DES redshift distribution. Inspecting the best-fit templates for those galaxies where the NIR data reduces the photometric redshift scatter, we find that quite often the NIR allows us to distinguish between an early-type SED and a reddened starburst SED. The larger scatter at the low and high redshift ends is driven mainly by the poorer sampling of the 4000\AA\@ break by the DES optical filters in these redshift ranges. At low redshifts, $u$-band photometry can help traverse the Balmer break while at higher redshifts, near infrared photometry is required. The minimum in the photometric redshift scatter occurs at $z\sim$0.7 where the 4000\AA\@ break is in the middle of the optical filters. 

Although the weighting method has been used to try and overcome biases associated with the available spectroscopic sample, in this single field, the statistics may also be strongly influenced by a small number of highly weighted galaxies. Nevertheless, these first results on the photometric redshifts from DES and VHS data are encouraging and suggest that the near infrared photometry derived from the DES detection images, can be used to improve the DES-only photometric redshift performance. In the future, widening the analysis to the full DES+VHS area will enable us to have much larger sub-samples of both red galaxies as well as $z \gtrsim 1$ galaxies, where the VHS data is expected to result in the most substantial improvements to the photometric redshift performance. We note that the RMS scatter in the photometric redshift at $z>1$ seen in the right panel of Figure \ref{fig:metrics}, is entirely consistent with the expectations from simulations where the VHS data was found to reduce the scatter from $\sim$0.2 to $\sim$0.15 at these high redshifts \citep{Banerji:08}.

In the low signal-to-noise regime probed by the VHS forced photometry catalogues, logarithmic magnitude estimates are also clearly biased and fluxes and associated errors provide a more unbiased measure of the galaxy photometry (see e.g. discussion in Appendix A of \citealt{Mortlock:12b}). Adapting existing photometric redshift codes to work with negative flux measurements and associated errors will therefore undoubtedly improve the constraints on photometric redshifts from the combined DES+VHS dataset. 

\section{TARGET SELECTION WITH DES+VHS}

\label{sec:target}

The next generation of wide-field spectroscopic surveys e.g. 4MOST \citep{deJong:12} and DESI \citep{Levi:13}, will target a wide variety of extragalactic sources in order to constrain both models of galaxy formation and cosmic acceleration. Targets for these spectroscopic surveys will need to be selected using photometric data e.g. from DES and VHS \citep{Jouvel:13, Schlegel:11}. There are primarily three types of targets for which spectroscopic redshifts will be obtained - Luminous Red Galaxies (LRGs) at $z < 1.2$, Emission Line Galaxies (ELGs) out to z$\sim$1.5 (depending on the red cut-off of the spectrograph) and quasars at $z\sim2-3$ where optical spectrographs can detect the Ly$\alpha$ forest. While the blue ELGs can be effectively selected using optical colours only, the near infrared VHS data should aid the selection of both LRGs and quasars for these next generation wide-field spectroscopic surveys. We investigate the utility of our DES+VHS catalogues for both LRG and quasar selection here. 

As described in Section \ref{sec:ozdes}, spectroscopic follow-up of DES targets in the SN fields is currently being conducted as part of the OzDES survey. OzDES can therefore serve as a pilot study for designing and honing photometric target selection methods using DES+VHS data, for the next generation of wide-field spectroscopic surveys. Once again we therefore make use of the OzDES redshift catalogue in order to assess the effectiveness of simple optical+NIR target selection criteria for isolating both high redshift LRGs and quasars. These are meant to serve as illustrative examples of the feasibility of using DES+VHS data for target selection for future surveys. In the future, these selection criteria will need to be further improved to produce the target densities necessary for different science applications.  

\subsection{Luminous Red Galaxies at High Redshift}

\begin{figure*}
\begin{center}
\includegraphics[scale=0.4,angle=0]{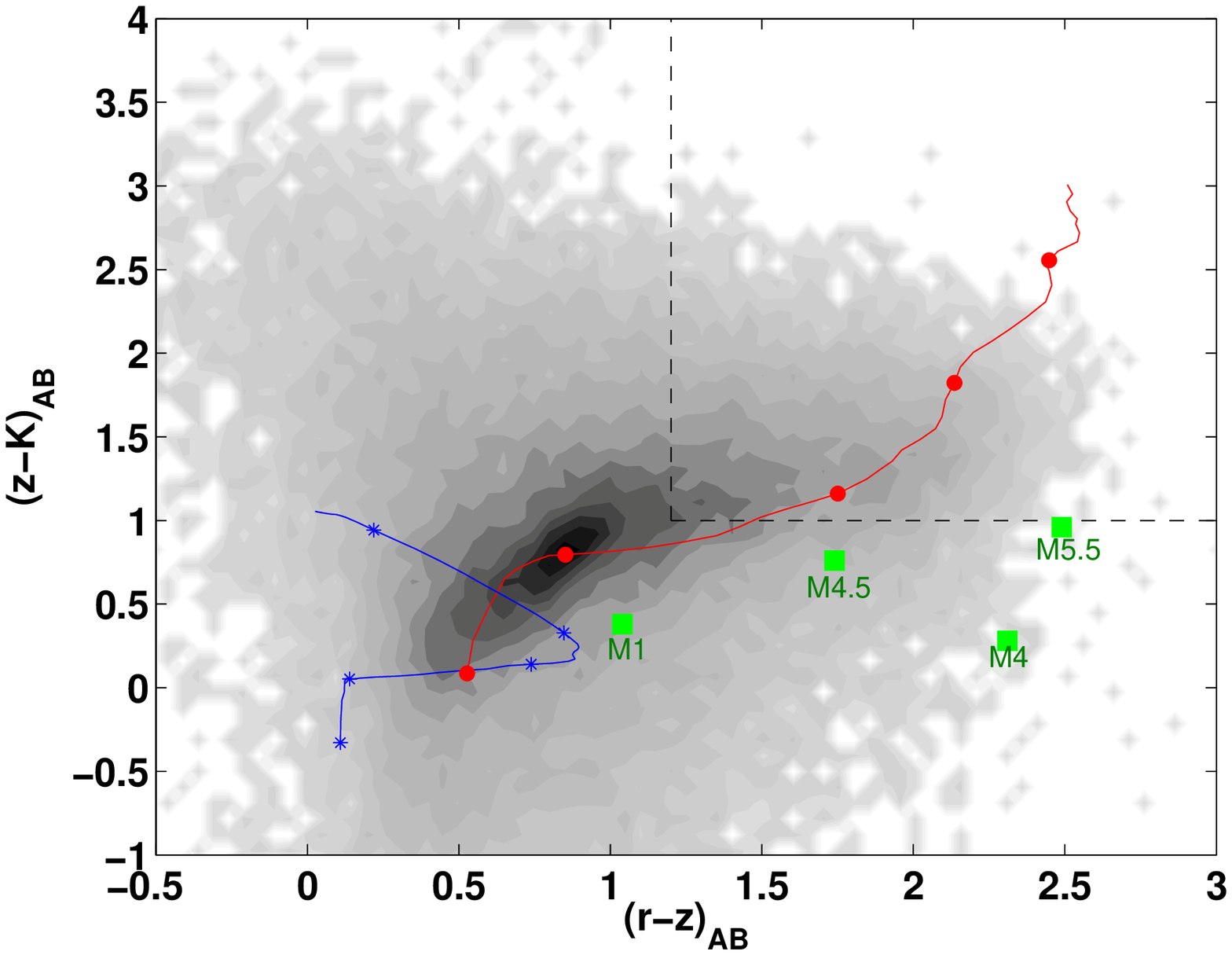}
\includegraphics[scale=0.5,angle=0]{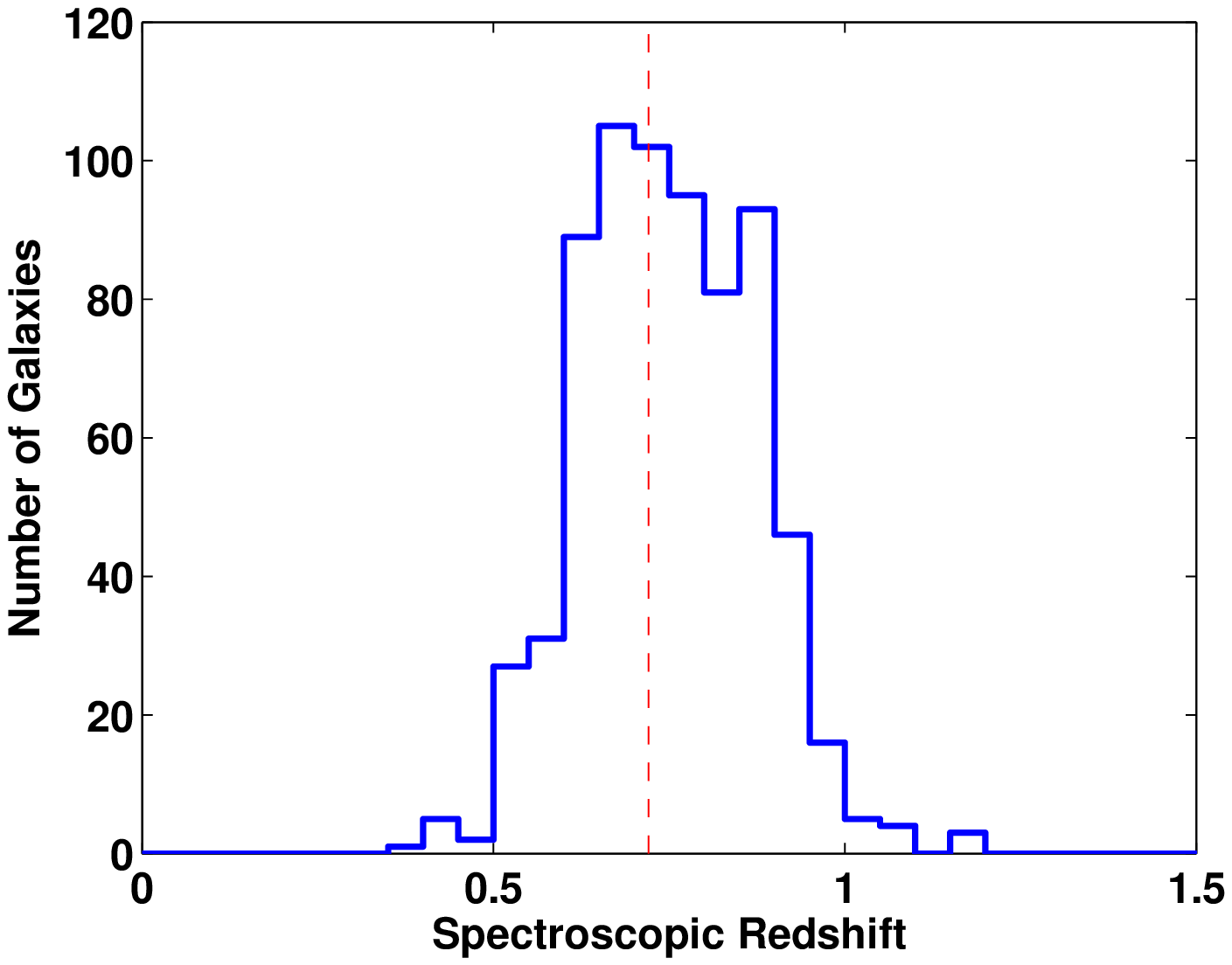}
\caption{\textit{Left:} Source density plot showing galaxies in the SN-X3 field in the DES+VISTA $(r-z)$ versus $(z-K)$ colour-colour plane. The red line represents the colour of a typical early-type galaxy template from \citet{Jouvel:09} in the redshift range $0<z<2$ while the blue line shows a late-type galaxy template from the same work. The green squares denote some representative colours of early-type M-stars from \citet{Leggett:00}, which have very similar optical colours to LRGs but are significantly bluer in terms of their infrared colour. Markers indicate redshift steps of 0.4 in these tracks. Our high-redshift LRG selection box for OzDES is represented by the top right of the plot. \textit{Right:} Redshift distribution of all LRG targets in the field that have secure redshift identifications. The median redshift of 0.72 is marked with a vertical line.}
\label{fig:lrgz}
\end{center}
\end{figure*}

We have employed a simple colour-selection in DES+VHS $rzK$ space to select Luminous Red Galaxies (LRGs) at high redshifts of $z\gtrsim$0.5 which extends the redshift range of currently available spectroscopic samples for photometric redshift calibration and large scale structure studies. The colour selection is shown in Figure \ref{fig:lrgz}, where we demonstrate that the use of the VHS NIR photometry is particularly useful for the removal of galactic M-stars, which have very similar optical colours to the LRGs.

\begin{equation}
\begin{split}
(r-z) > 1.2 \\
(z-K)>1.0 \\
i < 21.5 \\
i\_spread\_model > 0.0028
\end{split}
\label{eq:rzK}
\end{equation}

We apply this selection over the SN-X3 field which results in $\sim$2500 LRG candidates over 3 sq-deg. Out of these, there are 811 galaxies in total with spectroscopic redshift measurements from a variety of spectroscopic surveys in this field including AAOmega observations obtained as part of OzDES Yr 1 where the specific LRG selection described above was applied to select targets. Seven hundred and nine of the 811 have high confidence redshift measurements. The number of LRG targets with secure redshifts from OzDES only is 291, which represents $\sim$40\% of the secure redshift sample. The redshift distribution for all 709 galaxies is shown in Figure \ref{fig:lrgz}. Only 1 target has $z=3.33$ and is not an LRG and there are three stars that were selected as LRGs. The rest of our targets with secure redshifts, are all at 0.3$<z<1.2$ so for the subset of targets for which secure redshift identifications are possible, the contamination fraction to the colour-selection from non-LRGs is $<$0.5\%. We note that there may be a higher proportion of non-LRGs among those targets for which secure redshifts have not been measured and a thorough analysis of the purity and completeness of the LRG selection is deferred to future OzDES specific papers. 

We confirm through visual inspection of the spectra of the $0.3<z<1.2$ targets, that these are mostly early type galaxies with a prominent Balmer break. The median redshift is 0.72 and the median $K_s$-band magnitude is 18.9. We can see that the DES+VHS colour-selection is therefore effectively picking high redshift LRGs. Optical colour selection of LRGs for example in the BOSS survey, results in a sample with a lower median redshift of $\sim$0.5 \citep{White:11}, albeit using a smaller optical telescope, which is less sensitive at red wavelengths than DECam. However, at $z\sim0.7$, contamination from galactic M-stars can be an issue for optical only selection methods. Indeed \citet{Ross:08} find that optical only colour selection methods suffer from $\sim$16-30\% contamination from these M-stars and the NIR photometry helps to effectively alleviate this problem. As mentioned above, using the NIR selection, $<$0.5\% of spectroscopically confirmed LRG targets are found to be galactic M-stars. Similar target selection criteria can therefore be applied to the combined DES+VHS data resulting from the final 5 year survey in order to select extragalactic LRG targets for wide-field spectroscopic surveys like 4MOST and DESI. 

\subsection{Quasars}

\label{sec:agn}

Next-generation wide-field spectroscopic surveys such as DESI also aim to obtain spectra for quasars at $z\sim2-3$ in order to measure the Baryon Acoustic Oscillations (BAO) at high redshifts through the Ly$\alpha$ forest. Bright quasars illuminate the intergalactic medium at these redshifts enabling detailed measurements of the clustering properties of clumps of neutral hydrogen gas along the line of sight between us and the quasar, which show up as absorption features in the quasar spectra. Bright quasars therefore need to be isolated as targets for spectroscopic surveys from photometric catalogues like DES+VHS. 

Quasars can be distinguished from similarly blue star-forming galaxies through morphological classification. The brightest quasars suitable for Ly$\alpha$ forest measurements, appear as unresolved, nuclear point sources in imaging data and are easily separated from extended sources like galaxies. More difficult is the separation of quasars from stars which substantially outnumber the quasars in terms of their space density on the sky, particularly at bright magnitudes. The optical colours of $z\sim2-3$ quasars closely resemble those of stars and the best separation is seen in the $ugr$ colour-colour plane where the well-known UV-excess of quasars relative to stars, can be used as a discriminant between the two populations \citep{Richards:02}. However, in the southern hemisphere, DES lacks $u$-band coverage. Alternative selection criteria therefore need to be devised, for example using quasar variability to distinguish them from stars \citep{Schlegel:11}. 

It has been known for some time that in addition to the well-known UV excess in quasar spectral energy distributions, they also display an excess in the $K$-band relative to stars \citep{Warren:00}, a property that has been exploited to conduct wide-field quasar surveys using the near infrared \citep{Maddox:08, Maddox:12}. As shown in \citet{Maddox:12}, the KX quasars are actually more effective than the UVX method in selecting complete samples at $2<z<3$ where the optical quasar colours begin to run into the stellar locus. Our DES+VHS catalogues may therefore be useful for photometric quasar selection given the lack of $u$-band data in DES. \citet{Maddox:08} used a simple colour selection in the $gJK$ colour-colour plane to photometrically select quasars from the SDSS+UKIDSS. Here we present a very similar $giK$ selection with the added advantage that at fainter magnitudes it makes use of the higher signal-to-noise $i$-band data from DES. We also make use of \textit{WISE} photometry \citep{Wright:10} at 3.4$\mu$m (W1) and 4.6$\mu$m (W2) to further discriminate between quasars and stars. We note that the \textit{WISE} data contains quasars out to the highest redshifts \citep{Blain:13} and can therefore effectively be used to target bright quasars at all redshifts. The selection for quasars is defined as follows:

\begin{equation}
\begin{split}
(g-i)_{\rm{AB}} < 1.1529 \times (i_{\rm{AB}} - K_{\rm{Vega}}) - 1.401 \\
(W1-W2)>0.7 \\
-0.003<i\_spreadmodel<0.0028 \\
i<21.5
\end{split}
\label{eq:agn}
\end{equation}

Applying the DES+VHS colour and morphology selection to our optical+NIR catalogue in the SN-X3 field, we find $\sim$1800 $i<21.5$ candidates over the 3 sq-deg. These are then matched to \textit{WISE} data from the \textit{AllWISE} release with $\sim$75\% yielding \textit{WISE} counterparts. Applying the \textit{WISE} colour selection, we are left with 200 candidates over the 3 sq-deg. In order to test whether the selection has been effective in isolating luminous quasars, we once again make use of the OzDES master redshift catalogue which includes quasars from VVDS-Deep, VIPERS, spectroscopic follow-up of X-ray selected AGN from \citet{Stalin:10} as well as AAOmega observations of the field. Out of the 200 targets, 127 currently have high confidence redshift measurements. Fifty one of the 127 spectroscopic objects are quasars at $1.8<z<3.5$ and only 13 are stars representing a stellar contamination fraction of $\sim$10\% for those objects with secure redshifts. In addition, there are 5 galaxies and the remaining objects are all lower redshift quasars. Although the numbers are small over this single DECam pointing, we conclude that the DES+VHS catalogues together with \textit{WISE} data are proving effective in photometrically selecting quasars. These catalogues have also been used for target selection of quasars for reverberation mapping monitoring as part of the OzDES project and the larger quasar catalogue over the wider field-of-view, will be presented in Yuan et al. (in preparation). 

\section{OTHER APPLICATIONS}

In this first paper describing the optical + near infrared catalogues from DES and VHS, we have demonstrated that these catalogues can be effectively used to improve photometric redshifts from the DES-only data as well as for target selection for wide-field spectroscopic surveys. Through simple colour selection criteria that have been applied to the joint catalogues from the DES Science Verification period, we have shown that DES+VHS can be used for the selection of both broad-line quasars as well as Luminous Red Galaxies at high redshifts of $z\sim 0.7$. In future work, these selection criteria will be honed to provide purer and more complete samples of these populations. 

The current work has focussed in particular on photometric redshifts and target selection with DES+VHS data but we highlight that the joint photometry catalogue produced from the two surveys will also enable a range of other investigations which will significantly enhance the science possible from DES alone. We have shown in Section \ref{sec:sg} that the VHS near infrared colours could be useful for star-galaxy separation in DES. The inclusion of near infrared photometry will also result in improved stellar mass estimates for DES galaxies as the older long-lived stellar populations in galaxies that contribute most to their total mass, are redder in colour (e.g. \citealt{Bell:03, Drory:04}). Even for the bluest galaxies, near infrared photometry can help break degeneracies in spectral energy distribution fit parameters (e.g. \citealt{Banerji:13b}) demonstrating the utility of the VHS data in constraining stellar masses of DES galaxies. DES is also capable of potentially identifying very massive (M$_*>10^{12}$M$_\odot$) galaxies at $z > 4$, should they exist, and the near infrared data will be useful in distinguishing these galaxies from other contaminant populations \citep{Davies:13}. 

\subsection{Extremely Red Objects}

The VHS data will also be more sensitive than DES to rare populations of extremely red objects such as high redshift quasars \citep{Venemans:07, Mortlock:11, Willott:09}, cool dwarfs in our own Milky Way (e.g. \citealt{Lodieu:12, Burningham:13}) as well as the dustiest quasars which represent supermassive black-holes in the process of formation (e.g. \citealt{Banerji:12, Banerji:13a}). Many of these will be undetected or extremely faint in DES but the optical photometry is nevertheless necessary in order to verify the extreme colours of these rare sources and discriminate between them and other contaminant populations. 

Initial investigation of the colours of high redshift quasars in the DES+VHS Science Verification data shows that the forced photometered VHS fluxes still provide useful colour information for $z=6$ quasars. The quasar CFHQS J022743$-$060530 was discovered by the Canada France Hawaii Quasar Survey and lies at a redshift of $z=6.20$ \citep{Willott:09}. It is the only existing high redshift quasar to overlap the DES Science Verification observations. The search for and discovery of new high redshift quasars in the DES Science Verification data are presented in Reed et al. (in preparation). From the DES imaging, we find that CFHQS0227$-$0605 is undetected in the DES $i$-band and has a $z$-band PSF magnitude of 22.16. We measure colours in a 1 arc sec aperture of $(i-z)$=3.66 and $(z-J)$=0.42 where the latter comes from the forced photometry DES+VHS catalogue. We note here that unlike the forced photometry catalogues in the SN-X3 field presented in this paper, the joint DES+VHS forced photometry catalogue overlapping CFHQS J022743$-$060530 uses the $riz$ combined image for object detection and is therefore sensitive to objects that are not detected in the $r$-band. As mentioned in Section \ref{sec:des} however, these $r$-band drop-outs are extremely rare and use of the $r$-band for detection does not affect the conclusions of this paper.

CFHQS J022743$-$060530 is among the faintest $z=6$ quasars currently known and has a signal-to-noise of $\sim$1 in our $J$-band forced photometry catalogues. The $(z-J)$ colour is typically used to rule out a cool star classification as cool dwarfs in our own Milky Way are the main contaminants in high redshift quasar searches. Despite the low signal-to-noise, we find that the measured $(z-J)$ colour of this source is consistent with that of a high redshift quasar \citep{Willott:09}. A higher signal-to-noise detection in the $J$-band leading to a redder $(z-J)$ colour, would favour a cool star identification for this object. Even in the low signal-to-noise regime, limits on the near infrared photometry obtained using the DES detections can therefore help set important constraints on the colours for the identification of high redshift quasars. 

\section{CONCLUSIONS}

We have described the construction of a joint optical and near infrared ($grizYJK$) photometric catalogue from the Dark Energy Survey and VISTA Hemisphere Survey over a single 3 sq-deg DECam pointing centred at 02h26$-$04d36 targeted as part of the DES Science Verification period. The availability of ancillary multi-wavelength photometric data as well as large spectroscopic samples in this field, allows us to validate the quality of our photometric catalogues and assess their utility for a range of science applications. The VHS forced photometry catalogues are shown to be robust to within a few per cent to changes in the photometric processing pipeline and the errors on the photometry are dominated by the random sky noise. We demonstrate that using the deep DES detection images to extract fluxes from the VHS images at the DES positions results in a factor of $\sim$4.5 increase in the number of sources with optical+NIR photometry. While a simple catalogue match between DES and VHS provides $\sim$10,000/deg$^2$ galaxies with joint optical and NIR photometry, performing forced photometry allows us to increase this number to $\sim$45,000 galaxies/deg$^2$ albeit now with lower average signal-to-noise in the infrared. Almost 70\% of DES sources in this field have useful NIR flux measurements through the forced photometry. We show that the near infrared colours of galaxies in this forced photometry catalogue are consistent with templates generated from deep photometry in the COSMOS field. By matching our catalogue to deep near infrared data from the VISTA VIDEO survey, we show that the number counts agree very well with the VIDEO data over the magnitude range over which the two surveys overlap but the DES detections allow us to probe $\sim$1.5 mag deeper in the near infrared compared to the VHS 5$\sigma$ catalogues. In this paper we assess how useful this joint catalogue is for a range of science applications, concentrating in particular on photometric redshifts and target selection for future wide-field spectroscopic surveys. 

The addition of near infrared VHS data to DES photometry results in a modest improvement in the overall photometric redshift performance in particular reducing the scatter on the photometric redshift by $\sim$7-12\% (depending on the metric used to compute this scatter). We find that the VHS data allow us to quantify both the bias and scatter on the photometric redshift to better accuracy than with DES alone and note that there is some evidence that the VHS data are particularly helpful in reducing the photometric redshift scatter at $z<0.5$ and $z>1$. While these results are encouraging, we caution that they are currently affected by the small number statistics of available spectroscopic samples overlapping the single field where we have conducted our analysis and are also sensitive to the photometric redshift algorithm used. Our spectroscopic catalogues are over-represented at low redshifts and bright magnitudes where galaxy photometric redshifts are already well estimated with optical data. While we have attempted to correct for this bias by weighting the spectroscopic sample such that it has the same magnitude distribution as our photometric sample, in this single field, the results could be strongly influenced by a small number of highly weighted galaxies. Nevertheless, the photometric redshift analysis conducted here suggests that the VHS forced photometry is likely to improve the photometric redshift estimates from DES alone, particularly when larger samples with higher fractions of both high-redshift and intrinsically red galaxies become available. 

We introduce a simple DES+VHS colour selection scheme for the identification of high redshift Luminous Red Galaxies. Our proposed selection in the $rzK$ colour-colour plane gives $\sim$900 targets per sq-deg down to a magnitude limit of $i<21.5$. By matching these targets to existing spectroscopic databases as well as AAOmega observations of this field obtained as part of the OzDES project, we verify that the selection proves extremely effective in isolating high redshift LRGs with a median redshift of 0.72. Contamination from non-LRGs is estimated to be $<$0.5\% for those objects with secure redshifts although we note that the contamination may be higher for the subset of objects where secure redshift identifications were not possible. This contamination is considerably lower than found using optical-only colour selection methods and the additional VHS fluxes are shown to be useful for discriminating between LRGs and galactic M-stars.

For the purposes of quasar selection we propose the use of DES, VHS and \textit{WISE} catalogues. As DES will not have $u$-band observations, quasars cannot be separated from stars using the traditional UVX method which exploits the excess seen in quasar spectra relative to stars at UV wavelengths. Instead, we make use of the KX method and propose a simple $giK$ colour selection together with a \textit{WISE} colour cut for the identification of luminous quasars. Our proposed selection gives a target density of $\sim$70 bright quasars per sq-deg. Once again utilising the many spectroscopic datasets over this field, we show that a large fraction ($\sim$40\%) of the bright quasar targets with secure redshifts lie at $1.8<z<3.5$ and will therefore be suitable for Ly$\alpha$ forest measurements. Contamination from stars and low redshift galaxies is estimated to be $\sim$14\% for the subset of quasar targets with robust redshift measurements and most of the remaining sources with secure redshift identifications correspond to quasars at $z<1.8$. 

We conclude that the DES+VHS catalogues will form an important dataset for obtaining good photometric redshifts of galaxies in the southern hemisphere as well as aiding target selection for future wide-field spectroscopic surveys. The VHS data will also enable a range of other science investigations that will not be possible from DES alone e.g. improved stellar masses for DES galaxies, the identification of high redshift and highly reddened quasars and the discovery of cool, low-mass stars in our own Milky Way. Joint photometry from DES and VHS at the pixel level will eventually result in an optical+NIR catalogue of more than 200 million galaxies and $\sim$70 million point sources over 4500 sq-deg of the southern sky. Constructing these joint catalogues will therefore enhance the science possible from both surveys and contribute significantly to the landscape of wide-field photometric surveys in the southern hemisphere over the next decade. 

\section*{Acknowledgements}

We thank the referee, Nicholas Cross, for a very useful report on this manuscript. MB acknowledges a postdoctoral fellowship via OL's Advanced European Research Council Grant (TESTDE). 

Funding for the DES Projects has been provided by the U.S. Department of Energy, the U.S. National Science Foundation, the Ministry of Science and Education of Spain, 
the Science and Technology Facilities Council of the United Kingdom, the Higher Education Funding Council for England, the National Center for Supercomputing 
Applications at the University of Illinois at Urbana-Champaign, the Kavli Institute of Cosmological Physics at the University of Chicago, Financiadora de Estudos e Projetos, 
Funda{\c c}{\~a}o Carlos Chagas Filho de Amparo {\`a} Pesquisa do Estado do Rio de Janeiro, Conselho Nacional de Desenvolvimento Cient{\'i}fico e Tecnol{\'o}gico and 
the Minist{\'e}rio da Ci{\^e}ncia e Tecnologia, the Deutsche Forschungsgemeinschaft and the Collaborating Institutions in the Dark Energy Survey.

The Collaborating Institutions are Argonne National Laboratories, the University of California at Santa Cruz, the University of Cambridge, Centro de Investigaciones Energeticas, 
Medioambientales y Tecnologicas-Madrid, the University of Chicago, University College London, the DES-Brazil Consortium, the Eidgen{\"o}ssische Technische Hochschule (ETH) Z{\"u}rich, 
Fermi National Accelerator Laboratory, the University of Edinburgh, the University of Illinois at Urbana-Champaign, the Institut de Ciencies de l'Espai (IEEC/CSIC), 
the Institut de Fisica d'Altes Energies, the Lawrence Berkeley National Laboratory, the Ludwig-Maximilians Universit{\"a}t and the associated Excellence Cluster Universe, 
the University of Michigan, the National Optical Astronomy Observatory, the University of Nottingham, The Ohio State University, the University of Pennsylvania, the University of Portsmouth, 
SLAC National Laboratory, Stanford University, the University of Sussex, and Texas A\&M University.

The DES participants from Spanish institutions are partially supported by MINECO under grants AYA2009-13936,
AYA2012-39559, AYA2012-39620, and FPA2012-39684, which
include FEDER funds from the European Union. 

We are grateful for the extraordinary contributions of our CTIO colleagues and the DES Camera, Commissioning and Science 
Verification teams in achieving the excellent instrument and telescope conditions that have made this work possible.
The success of this project also relies critically on the expertise and dedication of the DES Data Management organisation.

The analysis presented here is based on observations obtained as part of the VISTA Hemisphere Survey, ESO
Progam, 179.A-2010 (PI: McMahon) and data products from observations made with ESO Telescopes at the La Silla Paranal Observatory under programme ID 179.A-2006 (PI: Jarvis). 

Data for the OzDES spectroscopic survey were obtained with the Anglo-Australian Telescope (program numbers 12B/11 and 13B/12). Parts of this research were conducted by the Australian Research Council Centre of Excellence for All-sky Astrophysics (CAASTRO), through project number CE110001020. TMD acknowledges the support of the Australian Research Council through Future Fellowship, FT100100595. 

\bibliography{}

\section*{Affiliations}
{\small
$^{1}$Department of Physics \& Astronomy, University College London, Gower Street, London WC1E 6BT, UK. \\
$^{2}$ Fermi National Accelerator Laboratory, P.O. Box 500, Batavia, IL 60510 USA \\
$^{3}$ Institute of Astronomy, University of Cambridge, Madingley Road, Cambridge CB3 0HA, UK. \\
$^{4}$ Kavli Institute for Cosmology, University of Cambridge, Madingley Road, Cambridge CB3 0HA, UK. \\
$^{5}$ Institut de Ci\`encies de l'Espai, IEEC-CSIC, Campus UAB, Facultat de Ci\`encies, Torre C5 par-2, 08193 Bellaterra, Barcelona, Spain. \\
$^{6}$ Institut d'Astrophysique de Paris, Univ. Pierre et Marie Curie \& CNRS UMR7095, F-75014 Paris, France. \\
$^{7}$ Observat\'orio Nacional, Rua Gal. Jos\'e Cristino 77, Rio de Janeiro, RJ - 20921-400, Brazil. \\
$^{8}$ Laborat\'orio Interinstitucional de e-Astronomia - LIneA, Rua Gal. Jos\'e Cristino 77, Rio de Janeiro, RJ - 20921-400, Brazil. \\
$^{9}$ Department of Astronomy, University of Illinois, Urbana, IL 61820, USA. \\
$^{10}$ Department of Physics, University of Michigan, Ann Arbor, Michigan 48109, USA. \\
$^{11}$Departamento de F\'{i}sica Matem\'{a}tica, Instituto de F\'{i}sica, Universidade de Sao Paulo, Sao Paulo, SP CP 66318, CEP 05314-970, Brazil\\
$^{12}$ Max-Planck-Institut f\"{u}r extraterrestrische Physik, Giessenbachstr.\ 85748 Garching, Germany\\
$^{13}$ Institut de F\'{i}sica d'Altes Energies, Universitat Aut\`{o}noma de Barcelona, E-08193 Bellaterra (Barcelona), Spain. \\
$^{14}$ Space Telescope Science Institute (STScI), 3700 San Martin Drive, Baltimore, MD 21218 USA. \\
$^{15}$ Department of Physics and Astronomy, University of Pennsylvania, Philadelphia, PA 19104, USA. \\
$^{16}$ Argonne National Laboratory, 9700 South Cass Avenue, Lemont, IL 60439, USA. \\
$^{17}$ Carnegie Observatories, 813 Santa Barbara St., Pasadena, CA 91101,USA. \\
$^{18}$ Institute of Cosmology \& Gravitation, University of Portsmouth, Portsmouth, PO1 3FX, UK. \\
$^{19}$ Research School of Astronomy and Astrophysics, The Australian National University, Canberra, ACT 2611, Australia. \\
$^{20}$ ARC Centre of Excellence for All-sky Astrophysics (CAASTRO). \\
$^{21}$ Kavli Institute for Particle Astrophysics and Cosmology 452 Lomita Mall, Stanford University, Stanford, CA, 94305. \\
$^{22}$ School of Mathematics and Physics, University of Queensland, QLD, 4072, Australia. \\
$^{23}$ George P. and Cynthia Woods Mitchell Institute for Fundamental Physics and Astronomy, and Department of Physics and Astronomy, Texas A \& M University, College Station, TX 77843-4242, USA. \\
$^{24}$ Department of Physics, Ludwig-Maximilians-Universit\"{a}t, Scheinerstr.\ 1, 81679, M\"{u}nchen, Germany. \\
$^{25}$ Excellence Cluster Universe, Boltzmannstr.\ 2, 85748 Garching, Germany. \\
$^{26}$ Centre for Astrophysics and Supercomputing, Swinburne University of Technology, Hawthorn, VIC 3122, Australia. \\
$^{27}$ Department of Physics, The Ohio State University, Columbus, OH 43210, USA. \\
$^{28}$ Oxford Astrophysics, Department of Physics, Keble Road, Oxford, OX1 3RH, UK. \\
$^{29}$ Physics Department, University of the Western Cape, Bellville, 7535, South Africa. \\
$^{30}$ Lawrence Berkeley National Laboratory, 1 Cyclotron Road, Berkeley, CA 94720, USA. \\
$^{31}$ Australian Astronomical Observatory, North Ryde, NSW 2113, Australia. \\
$^{32}$ICRA, Centro Brasileiro de Pesquisas F\'isicas, Rua Dr. Xavier Sigaud 150, CEP 22290-180, Rio de Janeiro, RJ, Brazil. \\
$^{33}$ Instituci\'o Catalana de Recerca i Estudis Avan\c{c}ats, E-08010 Barcelona, Spain. \\
$^{34}$ Centro de Investigaciones Energ\'eticas, Medioambientales y Tecnol\'ogicas (CIEMAT), Avda. Complutense 40, Madrid, Spain\\
$^{35}$ SLAC National Accelerator Laboratory, Menlo Park, CA 94025, USA. \\
$^{36}$ Research School of Astronomy and Astrophysics, The Australian National University, Canberra, ACT 2611, Australia. \\
$^{37}$ National Center for Supercomputing Applications, 1205 West Clark St., Urbana, IL, USA 61801. \\
$^{38}$ Department of Physics, University of Illinois, 1110 W. Green St., Urbana, IL 61801, USA. \\
$^{39}$ Jodrell Bank Centre for Astrophysics, University of Manchester, Manchester M13 9PL, UK. \\

}

\end{document}